\documentclass{aa}  
\pdfoutput=1 
\usepackage{graphicx}
\usepackage{pdfpages}
\usepackage{txfonts}
\usepackage[colorlinks=true,citecolor=blue,pagebackref=false,pdftex]{hyperref}
\usepackage{natbib}
\bibpunct{(}{)}{;}{a}{}{,}
\begin{document}

%==============================================================================================================
%%%%%%%%%%%%%%%%%%%%%%%%%%%%%%%%%%%%%%%%%%%%%%%%%%%%%%%%%%%%%%%%%%%%%%%%%%%%%%%%%%%%%%%%%%%%%%%%%%%%%%%%%%%%%%%
                                                %%%%%%%%%%%%%%%%%%%%% Title/Authors/Abstract %%%%%%%%%%%%%%%%%%%%%
%%%%%%%%%%%%%%%%%%%%%%%%%%%%%%%%%%%%%%%%%%%%%%%%%%%%%%%%%%%%%%%%%%%%%%%%%%%%%%%%%%%%%%%%%%%%%%%%%%%%%%%%%%%%%%%
   \title{Highly inclined and eccentric massive planets}
   
   \subtitle{II. Planet-planet interactions during the disc phase}

   \author{Sotiris Sotiriadis \inst{1}
          \and
          Anne-Sophie Libert  \inst{1} 
          \and
          Bertram Bitsch          \inst{2}
          \and
          Aur\'elien Crida        \inst{3,4}                    
          }
          
   \institute{
         naXys, Department of Mathematics, University of Namur, 8 Rempart de la Vierge, 5000 Namur, Belgium,\\ 
         \email{sotiris.sotiriadis@unamur.be}
         \and
         Lund Observatory, Department of Astronomy and Theoretical Physics, Lund University, 22100 Lund, Sweden
         \and
         Laboratoire Lagrange, Universit\'e C\^ote d'Azur, Observatoire de la C\^ote d'Azur, CNRS, Boulevard de l'Observatoire, CS34220, 06304 Nice cedex 4, France
         \and
         Institut Universitaire de France, 103 Boulevard Saint Michel, 75005 Paris, France
              }

   \date{Received ....................; accepted ...............}

   % ABSTRACT
   \abstract                   
    {Observational evidence indicates that the orbits of extrasolar planets are more various than the circular and coplanar ones of the Solar system. Planet-planet interactions during migration in the protoplanetary disc have been invoked to explain the formation of these eccentric and inclined orbits. However, our companion paper (Paper I) on the planet-disc interactions of highly inclined and eccentric massive planets has shown that the damping induced by the disc is significant for a massive planet, leading the planet back to the midplane with its eccentricity possibly increasing over time.
    }
    {We aim to investigate the influence of the eccentricity and inclination damping due to planet-disc interactions on the final configurations of the systems, generalizing previous studies on the combined action of the gas disc and planet-planet scattering during the disc phase. 
    } 
    {Instead of the simplistic $K$-prescription, our n-body simulations adopt the damping formulae for eccentricity and inclination provided by the hydrodynamical simulations of our companion paper. We follow the orbital evolution of $11000$~numerical experiments of three giant planets in the late stage of the gas disc, exploring different initial configurations, planetary mass ratios and disc masses. 
    }
    { The dynamical evolutions of the planetary systems are studied along the simulations, with a particular emphasis on the resonance captures and inclination-growth mechanisms. Most of the systems are found with small inclinations ($\le 10^{\circ}$) at the dispersal of the disc. Even though many systems enter an inclination-type resonance during the migration, the disc usually damps the inclinations on a short timescale. Although the majority of the multiple systems in our simulations are quasi-coplanar, $\sim5\%$ of them end up with high mutual inclinations ($\ge 10^{\circ}$). Half of these highly mutually inclined systems result from two- or three-body mean-motion resonance captures, the other half being produced by orbital instability and/or planet-planet scattering. When considering the long-term evolution over $100$ Myr, destabilization of the resonant systems is common, and the percentage of highly mutually inclined systems still evolving in resonance drops to $30\%$. Finally, the parameters of the final system configurations are in very good agreement with the semi-major axis and eccentricity distributions in the observations, showing that planet-planet interactions during the disc phase could have played an important role in sculpting planetary systems.}
    {}

    \keywords{planet and satellites: formation -- planet-disc interactions -- planets and satellites: dynamical evolution and stability}

   \maketitle
   \enlargethispage*{5pt}
%______________________________________________________________________________________________________________
%%%%%%%%%%%%%%%%%%%%%%%%%%%%%%%%%%%%%%%%%%%%%%%%%%%%%%%%%%%%%%%%%%%%%%%%%%%%%%%%%%%%%%%%%%%%%%%%%%%%%%%%%%%%%%%                 
%%%%%%%%%%%%%%%%%%%%%%%%%%%%%%%%%%%%%%%%%%%%%%%%%%%%%%%%%%%%%%%%%%%%%%%%%%%%%%%%%%%%%%%%%%%%%%%%%%%%%%%%%%%%%%%
%==============================================================================================================

\section{Introduction}
%==============================================================================================================
%%%%%%%%%%%%%%%%%%%%%%%%%%%%%%%%%%%%%%%%%%%%%%%%%%%%%%%%%%%%%%%%%%%%%%%%%%%%%%%%%%%%%%%%%%%%%%%%%%%%%%%%%%%%%%%         
                        %%%%%%%%%%%%%%%%%%%%% Introduction %%%%%%%%%%%%%%%%%%%%%
%%%%%%%%%%%%%%%%%%%%%%%%%%%%%%%%%%%%%%%%%%%%%%%%%%%%%%%%%%%%%%%%%%%%%%%%%%%%%%%%%%%%%%%%%%%%%%%%%%%%%%%%%%%%%%%
         The surprising diversity of the orbital parameters of exoplanets shows that the architecture of many extrasolar systems is remarkably different from that of the Solar system and sets new constraints on planet formation theories (see \citealt{Winn2014} for a review). The broader eccentricity distribution of the detected planets, the existence of giant planets very close to their parent star (``hot Jupiters'') and the strong spin-orbit misalignment of a significant fraction ($\sim \!40\% $) of them \citep{Albrecht1} are several characteristics that appear to be at odds with the formation of the Solar system. Furthermore, some evidence of mutually inclined planetary orbits exists; for instance, the planets $\upsilon$ Andromedae $c$ and $d$ whose mutual inclination is estimated to $30^{\circ}$ \citep{Deitrick}.  

         It seems that the interactions between the planets and their natal protoplanetary disc, at the early stages of formation, play an important role in sculpting a planetary system. The angular momentum exchange between the newborn planets (or planetary embryos) and the gaseous disc tends to shrink the semi-major axis and keeps them on coplanar and near-circular orbits \citep{Cresswell2007,Bitsch2010,Bitsch2011}. If the planets are massive enough to carve a gap and push the material away from their orbit, they migrate, generally inwards, with a timescale comparable to the viscous accretion rate of the gas towards the star, a phenomenon referred to as \textit{Type-II} migration (\citealt{Lin1,Kley1,Nelson1}, and \citealt{Baruteau2014} for a review). The gas also affects the eccentricities and inclinations of the embedded planets \citep{GT1,Xiang2013,Bitsch2013}.
         
    Several authors, using analytical prescriptions, have investigated the impact of the planet-disc interactions on the orbital arrangement of planetary systems through three-dimensional (3D) n-body simulations. Studies in the context of Type-II migration have been performed for two-planet systems \citep{Thommes2003,Libert2009,Teyssandier2014} and three-planet systems \citep{Libert2011b}. A notable outcome of these works is that, during convergent migration, the system can enter an inclination-type resonance that pumps the mutual inclination of the planetary orbits. 

         After the dispersal of the disc, dynamical instabilities among the planets can result in planet-planet scattering and a subsequent excitation of eccentricities and inclinations \citep{Marzari1996,Rasio1996,Lin1997}. Such a mechanism has been proposed to explain the eccentricity distribution of giant extrasolar planets \citep{Marzari2002,Chatterjee2008,Juric,Ford2008,Petrovich2014}. The planet-planet scattering model is based on unstable initial conditions that do not take into account the imprint of the disc era \citep{Lega2013}.     
         
        More realistic approaches study the combined action of both previous mechanisms (Type-II migration and scattering). Parametric explorations via n-body simulations \citep{Adams2003,Moorhead2005,Matsumura2010,Libert2011a} have shown that dynamical instabilities can occur during the disc phase.  
        
        In n-body simulations, a damping prescription depending on a scaling factor is commonly used to mimic the influence of the disc on the eccentricities of the planets, which is usually a first-order approximation for the eccentricity damping timescale ($ \dot{e}/e = - K |\,\dot{a}/a|$). The considered $K$-factor, currently not well determined but usually estimated in the range $1-100$, determines the final eccentricity and inclination distributions (see discussion in Section \ref{discuss}). Moreover, no damping on the planet inclination is generally included in these studies (e.g. \citealt{Thommes2003,Libert2009,Libert2011a}). On the other hand, hydrodynamical simulations accurately modelize planet-disc interactions. Several works have investigated the evolution of the eccentricities of giant planets in discs using two-dimensional (2D) hydrodynamical codes (e.g. \citealt{Marzari2010,Moeckel2012,Lega2013}). The influence of the disc on planetary inclinations has been studied in our accompanying paper (\citealt{Bitsch2013}, hereafter denoted Paper I), where we performed 3D hydrodynamical numerical simulations of protoplanetary discs with embedded high mass planets, above 1 $M_{\rm Jup}$, computing the averaged torques acting on the planet over every orbit. As there is a prohibitive computational cost  to following the orbital evolution for timescales comparable with the disc's lifetime ($\sim 1-10$ Myr) and, as a consequence, to generating large statistical studies, we have derived an explicit formula for eccentricity and inclination damping, suitable for n-body simulations, as a function of eccentricity, inclination, planetary mass, and disc mass. 
        
        In this paper, we aim to improve the previous n-body studies combining planet-disc interactions and planet-planet interactions by adopting the damping formulae of Paper I. In other words, we combine the speed and efficiency of an n-body integration with a symplectic scheme, and an improved modelization of the gas effect promoted by hydrodynamical simulations. More specifically, we extend the work of \citet{Libert2011a}, focusing on the orbital evolution of three giant planets in the late stage of the gas disc. In particular, we add inclination damping, which was not taken into account in their previous work; thus, comparing our results with theirs allows us to highlight the role of this phenomenon. We have performed $11000$ numerical simulations, exploring different initial configurations, planetary mass ratios and disc masses. Our goal is to investigate the influence of the eccentricity and inclination damping due to planet-disc interactions on the final configurations of planetary systems and their inclination distribution. Through our parametric analysis, we also discuss the most common three-body resonance captures occuring during the migration of the planets, as well as the most frequent mechanisms producing inclination increase.    

        The paper is organized as follows. In Section~\ref{section2}, we describe the set-up of our numerical experiments. Typical dynamical evolutions of planetary systems during the disc phase are analyzed in detail in Section~\ref{section3}, while the orbital parameters and resonance configurations of our set of planetary systems at the dispersal of the disc are presented in Section~\ref{section4}. In Section~\ref{section5}, we address the question of the stability of the planetary systems formed in our simulations, by analyzing the effect of the long-term evolution on the architecture of the systems. Finally, our conclusions are given in Section~\ref{section6}.
%______________________________________________________________________________________________________________
%%%%%%%%%%%%%%%%%%%%%%%%%%%%%%%%%%%%%%%%%%%%%%%%%%%%%%%%%%%%%%%%%%%%%%%%%%%%%%%%%%%%%%%%%%%%%%%%%%%%%%%%%%%%%%% 
%%%%%%%%%%%%%%%%%%%%%%%%%%%%%%%%%%%%%%%%%%%%%%%%%%%%%%%%%%%%%%%%%%%%%%%%%%%%%%%%%%%%%%%%%%%%%%%%%%%%%%%%%%%%%%%
%==============================================================================================================

\section{Set-up of the n-body simulations}\label{section2}
%==============================================================================================================
%%%%%%%%%%%%%%%%%%%%%%%%%%%%%%%%%%%%%%%%%%%%%%%%%%%%%%%%%%%%%%%%%%%%%%%%%%%%%%%%%%%%%%%%%%%%%%%%%%%%%%%%%%%%%%% 
                                                %%%%%%%%%%%%%%%%%%%%% Physical modelling %%%%%%%%%%%%%%%%%%%%%
%%%%%%%%%%%%%%%%%%%%%%%%%%%%%%%%%%%%%%%%%%%%%%%%%%%%%%%%%%%%%%%%%%%%%%%%%%%%%%%%%%%%%%%%%%%%%%%%%%%%%%%%%%%%%%%
\subsection{Physical modeling}
        In this work, we analyze the evolution of three-planet systems embedded in a protoplanetary disc and evolving around a Solar-mass star. We consider systems with three fully formed gaseous giant planets that do not accrete any more material from the disc as they migrate inwards, towards the star. We use the symplectic integrator SyMBA \citep{symba}, which allows us to handle close encounters between the bodies by using a multiple time-step technique. Moreover, due to the interaction with the disc and the subsequent migration, some systems end up in a compact stable configuration very close to the star. In these systems, planets might have either low or high eccentricity values. For this reason, we adopt a symplectic algorithm that also has the desirable property of being able to integrate close perihelion passages with the parent star \citep{Levison1}. In order to ensure high resolution for close-in orbits, we set the time-step of our simulations at $dt = 0.001$ yr.

        The migration of the planets in Type-II migration, due to angular momentum exchange with the disc, is on a similar timescale as the viscous accretion time, $ t_\nu \sim r_{\rm pl}^{2} / \nu  $, where $\nu$ is the kinematic viscosity and $r_{\rm pl}$ is the radius of the planetary orbit. However, when the mass of the planet is comparable to the mass of the material in its vicinity, the migration rate scales with the ratio of the planetary mass over the local disc mass \citep{Ivanov99,Nelson1,Crida2007}. Thus, the rate for Type-II regime consists of two different cases, the disc-dominated case and the planet-dominated case:  
    \begin{equation}\label{tII}
          \tau_{II} = \frac{2}{3} \alpha^{-1} h^{-2} \Omega_{pl}^{-1} \times \max\left(1,\frac{M_{p}}{(4\pi/3)\Sigma(r_{\rm pl})r_{\rm pl}^{2}}\right), 
    \end{equation}
        where $\alpha$ is the Shakura-Sunyaev viscosity parameter \citep{Shakura1973}, $h$ the disc aspect ratio, $\Omega$ the orbital frequency of the planet and ${(4\pi/3)\Sigma(r_{\rm pl})r_{\rm pl}^{2}}$ is the local mass of the disc. The rate of Type-II is still under debate and may depart from the viscous time \citep{Has2013,Duffell2014,Durmann2015}. However, the scope of the present paper is not to study the Type-II regime (which would require hydrodynamical simulations), but planet-planet interactions during planetary migration. \citet{Teyssandier2014} have pointed out that the eccentricity and inclination evolution (which is of interest for us here) of migrating giant planets in the 2:1 MMR is not affected by the Type-II timescale. Therefore, we always use Eq.~\ref{tII} (like most previous similar studies), independent of $e$ and $i$ of the planets.
        
        The disc parameters are set to the classical values $\alpha=0.005$ and $h=0.05$. Four values of disc mass are considered in this study, being $4, 8, 16$ and $32 M_{\rm Jup}$, in order to verify the robustness of our results. The initial surface density profile is $\Sigma \propto r^{-0.5} $ and the disc's inner and outer edges are set to $R_{\rm in}=0.05$ AU and $R_{\rm out}=30$ AU. We apply a smooth transition in the gas-free inner cavity, following \citet{Matsumoto2012}, by using a hyperbolic tangent function, $\tanh{\frac{(r - R_{\rm in})}{\Delta r}}$, where $\Delta r=0.001$ AU. Following \citet{Libert2011a}, we apply an inward migration to the outer planet only. We start the evolution of the system with fully formed planets and consider that there is not enough gas to drive migration between them\footnote{When all the planets migrate and do not share a common gap, the rate for Type-II regime given by Eq. \ref{tII} leads to divergent migration (instead of convergent migration like in the present work) and the same phenomenons of resonance capture and eccentricity/inclination excitation are only temporarily observed (see e.g. \citet{Libert2011b}).}. To fully mimic the interactions with the disc, we use the damping formulae of Paper~I for eccentricity and inclination (see Sections \ref{formules} and \ref{discuss} for their full description).

%^^^^^^^^^^^^^^^^^^^^^^^^^^^^^^ %
%^^^^^^^^^^ Figure 1 ^^^^^^^^^^ %
%^^^^^^^^^^^^^^^^^^^^^^^^^^^^^^ %
 \begin{figure}%[t!]     %wste na "katsei" oso pio konta ginetai
       \centering
       \includegraphics[width=\hsize]{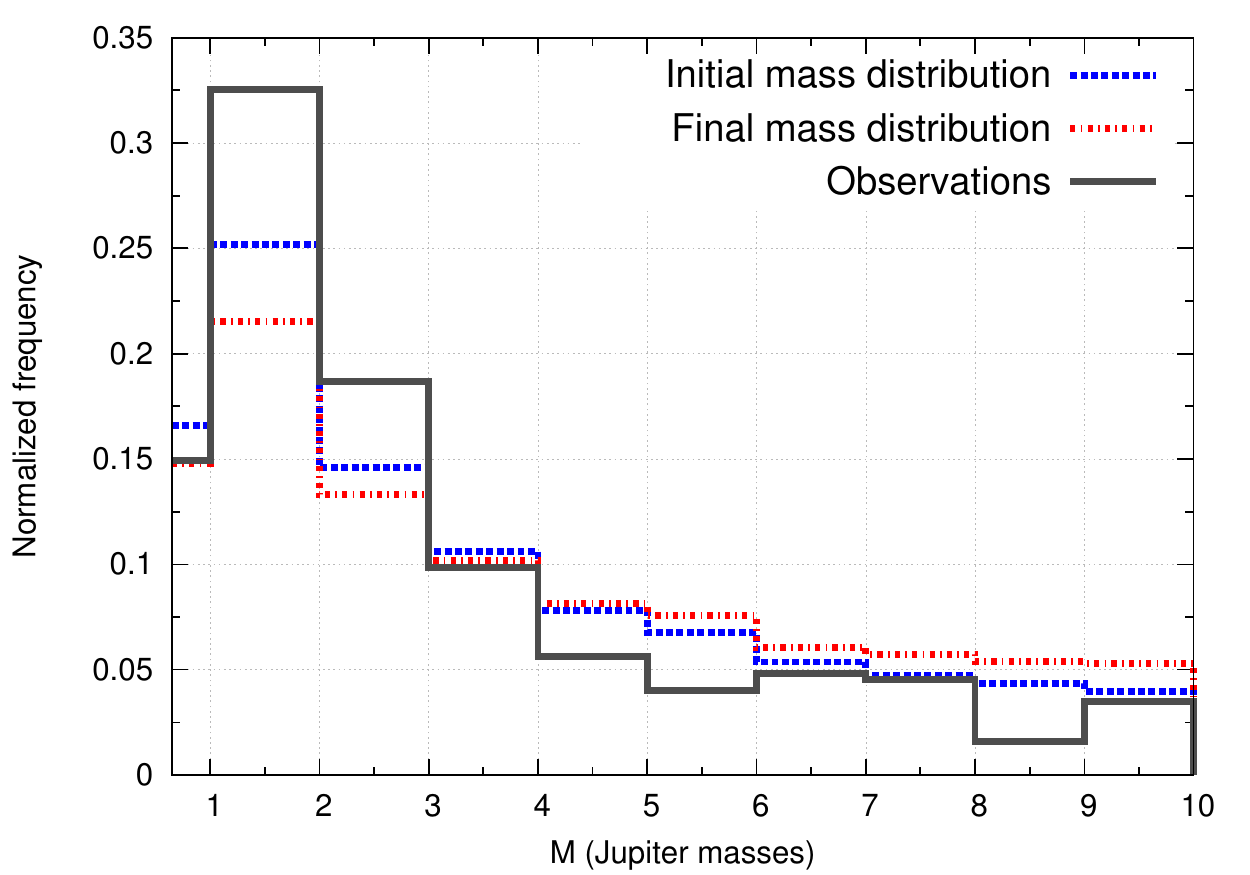} % Initial mass distribution
       \caption{The normalized initial and final mass distributions of our simulations compared with the mass-distribution (in the range $[0.65,10]$ $M_{\rm Jup}$) of the observed giant planets with $a>0.1$.}
       \label{fig:fig1}
 \end{figure}
%^^^^^^^^^^^^^^^^^^^^^^^^^^^^^^ %
%^^^^^^^^^^^^^^^^^^^^^^^^^^^^^^ %
        
        The effects of orbital migration, eccentricity, and inclination damping are added to the acceleration of the planets, as in \citet{Papaloizou_Larwood}:
        \begin{equation}
               \vec{a}_{\rm disc} = - \frac{\vec{v}}{\tau_{II}}~ - 2 \frac{(\vec{v}\cdot\vec{r})\vec{r}}{r^{2}\tau_{\rm ecc}} ~ - 2 \frac{(\vec{v}\cdot\vec{k})\vec{k}}{\tau_{\rm inc}}  ~,
        \end{equation}
        where $\tau_{\rm ecc}$ and $\tau_{\rm inc}$ denote the timescales for eccentricity and inclination damping, respectively (see Eqs. \ref{eq_e} and \ref{eq_i}). These rates are computed in every step of the integration and depend on the local surface density of the disc, the mass of the planet, and the eccentricity and inclination of the planet at the considered time. The damping forces were implemented in SyMBA in the same way as in \citet{Lee_Peale} (see their appendix for a detailed description of the method), such that it keeps the symmetry of the symplectic algorithm.      
        
        The decrease of the gas disc is implemented in two ways. In the {\it constant-mass model} (hereafter denoted CM), the mass of the circumstellar disc is kept constant for $0.8$ Myr, then the gas is instantly removed. We use this approach to understand, through statistical analyses, whether the inclination damping has a strong impact on the final configurations of planetary systems. A more realistic modelization of protoplanetary disc is the {\it decreasing-mass model} (DM), where we decrease its mass exponentially through the evolution of the system, with a dispersal time of $\sim 1$ Myr (e.g., \citet{Mama2009}). More specifically, we neglect the interaction with the disc when $dM_{\rm Disc}/dt < 10^{-9} M_{\rm star}/yr $. As a consequence, the surface density also evolves with time in the DM model.   
                
        During the evolution of a system, close encounters between the bodies can lead to planet-planet collisions, ejections from the system and collisions with the parent star. We treat a possible merge between two bodies as a totally plastic collision, when their distance becomes less than the sum of the radii of the planets. We assume Jupiter's density here when computing the radius from the mass. The boundary value for planet accretion onto the star is $0.02$ AU and the one for ejection from the system, $100$ AU.

%^^^^^^^^^^^^^^^^^^^^^^^^^^^^^^ %       
%^^^^^^^^^^ Table 1 ^^^^^^^^^^^ %
%^^^^^^^^^^^^^^^^^^^^^^^^^^^^^^ %
\begin{table}  % Constant mass simulations table
\begin{center}

  \caption{Initial disc mass, integration time, and number of simulations for each disc modelization.} \label{tab:title1} 
  \begin{tabular}{ r | r | r | r }

   \hline
 Model &   $M_{\rm Disc}$ $[M_J]$       & $N_{Systems}$          & $t_{stop}$ $[yr]$  \\ %\hline
 CM  & 4                                        & 600                            & $1.2\times10^6$               \\ %\hline
    &  8                                        & 600                            & $1.2\times10^6$               \\ %\hline
    & 16                                        & 2000                       & $1.2\times10^6$           \\ %\hline
    & 32                                        & 800                            & $1.2\times10^6$       \\ \hline    
 DM & 4                                 & 1000                           & $1.4\times10^6$               \\ %\hline
   & 8                                  & 1000                           & $1.4\times10^6$       \\ %\hline
   & 16                                 & 2400                       & $1.4\times10^6$         \\ %\hline
   & 32                                         & 1400                           & $1.4\times10^6$       \\
 & 8                                    & 400                            & $1.0\times10^8$               \\ %\hline    
   & 16                                         & 800                            & $1.0\times10^8$               \\ %\hline    
  \hline  
    \end{tabular}
   
\end{center}
\end{table}
%^^^^^^^^^^^^^^^^^^^^^^^^^^^^^^ %
%^^^^^^^^^^^^^^^^^^^^^^^^^^^^^^ %

        In our simulations, gas giant planets are initially on coplanar and quasi-circular orbits ($e \in [0.001,0.01]$ and $i \in [0.01^{\circ},0.1^{\circ}]$). The initial semi-major axis of the inner body is fixed to $3$ AU and the middle and outer ones follow uniform distributions in the intervals $a_{2}\in[4.3,6.5]$ AU and $a_{3}\in[8,16]$ AU, respectively. These initial semi-major axis distributions provide a wide range of possible resonance captures. We choose randomly initial planetary masses from a log-uniform distribution that approximately fits with the observational data\footnote{http://exoplanetarchive.ipac.caltech.edu/} in the interval $[0.65,10]$~$M_{\rm Jup}$ (black and blue curves in Fig. \ref{fig:fig1}). In Table \ref{tab:title1}, we describe the different sets of simulations we performed for both disc modelizations and different disc masses. In total, $4000$ simulations were realized for the CM model and $7000$ for the DM model. The computational effort required for our investigation was $\sim 1.8\times10^{5}$ computational hours.     

\subsection{Eccentricity and inclination damping formulae}\label{formules}
      The damping rates for eccentricity and inclination used in the present work originate from the accompanying Paper I, where the evolutions of the eccentricity and inclination of massive planets ($\ge$~$1$~$M_{\rm Jup}$) in isothermal protoplanetary discs have been studied with the explicit/implicit hydrodynamical code NIRVANA in 3D. The forces from the disc, acting onto the planet kept on a fixed orbit, were calculated, and a change of ${de}/{dt}$ and $di/dt$ was thereby determined.
      
      Eccentricity and inclination are mostly damped by the interactions with the disc. For highly inclined massive planets, the damping of $i$ occurs on smaller timescales than the damping of~$e$. The only exception is low-inclination planets with a sufficient mass ($\ge 4-5 M_{\rm Jup}$), for which the interactions of the planet with the disc result in an increase of the planet eccentricity. As a result, for single-planet systems, the dynamics tend to damp the planet towards the midplane of the disc, in a circular orbit in the case of a low-mass planet and in an orbit whose eccentricity increases over time due to the interactions
with the disc in the case of a high-mass planet and a sufficiently massive gas disc. The increase of the planetary eccentricity for massive planets may appear surprising, but it is a well-established phenomenon, further discussed in Section \ref{discuss}.
      
      The damping formulae depend on the planet mass $M_P$ (in Jupiter masses), the eccentricity $e_P$ and the inclination $i_P$ (in degrees). The eccentricity damping function is given by
      \begin{align}\label{eq_e}
      \frac{de}{dt} (M_P,e_P,i_P)=&-\frac{M_{disc}}{0.01\,M_\star}\left(a(i_P+i_D)^{-2b}+ci_P^{-2d}\right)^{-1/2} \nonumber \\
      & + 12.65 \, \frac{M_P M_{disc}}{M_{\star}^2}\, e_P \, \exp\left(-\left(\frac{(i_P/1^\circ)}{{M}_p}\right)^2\right) \ ,
      \end{align} 
      where $i_D={M}_p/3$ degrees, $M_{\star}$ is the mass of the star in solar mass, and with coefficients
      \begin{align}
      a_e(M_P,e_P) &=80\,e_P^{-2}\, \exp \left(-e_P^2 {M}_p / 0.26 \right) 15^{{M}_p} \, \left(20 + 11{M}_p-{M}_p^2 \right) \nonumber \\
        b_e(M_P) &= 0.3{M}_p \nonumber \\
        c_e(M_P) &= 450+2^{{M}_p} \nonumber \\
        d_e(M_P) &= -1.4+\sqrt{{M}_p}/6 \nonumber \ .
      %\label{eq:F_e}
      \end{align}
      The damping function for inclination is given, in degrees per orbit, by
      \begin{align}\label{eq_i}
      \frac{di}{dt}(M_P,e_P,i_P) &= -\frac{M_{disc}}{0.01\,M_\star}\left[ a_i\, \left(\frac{i_P}{1^\circ}\right)^{-2b_i}\exp(-(i_P/g_i)^2/2) \right.\\
      &   \hspace{3.5cm} + \left.c_i\,\left(\frac{i_P}{40^\circ}\right)^{-2d_i}\ \right]^{-1/2}, \nonumber
      \end{align}
      with coefficients
      \begin{align}
      a_i(M_P,e_P) &= 1.5 \cdot 10^4 (2-3e_P){{M}_p}^3\nonumber \\
      b_i(M_P,e_P) &= 1+{M}_p e_P^2 /10 \nonumber \\
      c_i(M_P,e_P) &= 1.2 \cdot 10^{6}/\big[ (2-3e_P) (5+e_P^2 ({M}_p+2)^3) \big] \nonumber \\
      d_i(e_P) &= -3+2e_P \nonumber \\
      g_i(M_P,e_P) &= \sqrt{3 {M}_p / (e_P+0.001)}\times 1^\circ. \nonumber 
      \end{align}

      Let us highlight the limitations of the implemented formulae. These expressions have been derived from fitting the results of hydrodynamical simulations for planets between $1$ and $10$ $M_{\rm Jup}$  (to cover the range of giant planets) with an eccentricity smaller than $0.65$ (the expression for coefficient $c_i$ is clearly not valid for $e_P>2/3$). Practically, when the eccentricity of a planet exceeds the limit value during the integration, the square root of a negative number must be computed, so we instead give a small number ($10^{-5}$) to the problematic factor ($2-3e_P$) of the inclination damping formula. Moreover, systems with an overly massive planet ($>10$~$M_{\rm Jup}$), and following a merging event,  will be disregarded from our parametric analysis. 
      
    An additional limitation concerning the planetary inclinations is that it is rather unclear if an inclined planet will continue to migrate on a viscous accretion timescale. \citet{Thommes2003} stated that the interaction of a planet away from the midplane with the gas disc is highly uncertain. Furthermore, it has been shown that Kozai oscillations with the disc govern the long-term evolution of highly inclined planets (e.g., \citet{Ajmia,Teyssandier2013658}; Paper I), which makes it hard to predict the exact movement of the planet and its orbital parameters at the dispersal of the disc. We note that the study of the exact Type-II timescales due to planet-disc interactions is beyond the scope of the present work; therefore, for simplicity and despite these limitations, the viscous accretion timescale is adopted independently of the planetary inclination in our simulations. However, as we will see in Sections \ref{4.3} and \ref{4.6}, reaching such high inclination during the disc phase is rather occasional, starting from a coplanar system in interaction with the disc. We note also that no inclination damping is applied for planets with inclinations below $0.5^{\circ}$.   
     
    In Eqs. \ref{eq_e} and \ref{eq_i}, the damping (and excitation) formulae scale with the mass of the disc. It should be noted that the total mass of the disc is actually an irrelevant parameter\,: whatever the power law chosen, the integration of $\Sigma(r)$ from $r=0$ to $+\infty$ diverges. Hence, the total mass in the grid actually depends at least as much on the boundaries as on the local surface density. But the damping is done locally by the disc in the neighborhood of the planet. Therefore, the $M_{\rm disc}$ parameter in Eqs. \ref{eq_e} and \ref{eq_i} is actually the disc mass between $0.2$ and $2.5\,a_{\rm Jup}$. It should be noted that this was not the total mass of gas included in the grid, because the grid was extended to $4.2\,a_{\rm Jup}$. In the present paper, for convenience, the notation $M_{\rm Disc}$ refers to the total mass of the disc, instead of the local one. Therefore, we have scaled the damping formulae accordingly, taking $M_{\rm disc}=M_{\rm Disc}\times (2.5/4.2)^{3/2}=0.4\, M_{\rm Disc}$.
     
    The previous formulae have been derived from hydrodynamical simulations with fixed disc parameters; for instance, the shape of the density profile, the viscosity, and the aspect ratio. A change of values of these parameters in the present study would not be consistent. We do not aim to realize an exhaustive study of the free parameters of the hydrodynamical simulations, but, with realistic parameters, analyze in detail the dynamical interactions between the planets in the disc. 

    The final state of single-planet systems in a protoplanetary disc has been studied in detail in Paper I. In the present work, we consider systems of multiple planets that excite each other’s inclination during the migration in the gas disc while suffering from the strong damping influence of the disc. When two planets share a common gap (i.e., mutual distance~<~$4\,~R_{\rm Hill}$), we reduce the effect of the damping by a factor of $2$, because each planet only interacts with either the inner or the outer disc, that is half the mass of gas a single planet would interact with. For reference, this applies for a long time scale in only $2\%$ of our simulations, since the planets are initially on well-separated orbits and never evolve in a resonance with close enough orbital separation to share a common gap according to our criterion. When three planets share a gap (which applies for a long time scale in only $<1\%$ of our simulations), the middle one is hardly in contact with the gas disc; therefore, we reduce the damping rates by a factor of 100 for the middle planet. Choosing 100 is relatively arbitrary, but the basic idea is that the middle planet's eccentricity and inclination are almost not damped by the inner and outer disc because they are too far away, and the gas density inside the gap is at most a hundredth of the unperturbed surface density. Taking 50 or 500 would not 
significantly change the evolution of the system: the middle planet is excited freely by the two others. 

\subsection{Discussion on damping timescales}\label{discuss}
        We stress that the final results of our simulations depend on the recipe chosen for the damping of the eccentricity and inclination of the planets. Here we review the benefits and limitations of the $K$-prescription and the prescription of Paper~I.

        In n-body simulations, a $K$-prescription for the eccentricity damping timescale,
\begin{equation}
\frac{\dot{e}}{e} = - K \left|\,\frac{\dot{a}}{a}\right|
,\end{equation}
is commonly used to mimic the influence of the disc on the eccentricities of the planets. For instance, \citet{Lee_Peale} reproduced the GJ\,876 system using a fixed ratio $K=100$ when migration and eccentricity damping are applied to the outer planet only, and $K=10$ when applied to the two planets. Concerning the formation of inclined systems, \cite{Thommes2003}, \cite{Libert2009}, and \cite{Teyssandier2014} performed extensive parametric studies for different $K$ values and orbital parameters of the planets. They showed that planets captured in a mean-motion resonance during Type-II migration can undergo an inclination-type resonance, if eccentricity damping is not too efficient. \cite{Libert2011a} and \cite{Libert2011b} generalized these results to three-planet systems, studying the establishment of three-planet resonances (similar to the Laplace resonance in the Galilean satellites) and their effects on the mutual inclinations of the orbital planes of the planets. These works concluded that the higher the value of $K$, the smaller the final eccentricities and the less the number of mutually inclined systems.

        However, since the actual value of $K$ is estimated in the wide range $1-100$, no eccentricity or inclination distributions can be inferred from these studies, the final system configurations being strongly dependent on the value of $K$ considered in the simulations. Moreover, the $K$-prescription approach is too simplistic, as highlighted in several works. For instance, \cite{autocrida} showed that for gap opening planets, the ratio K is not constant, but depends on the eccentricity of the planet (see their Fig. 3, fourth panel). Note also that the $K$-prescription has been developed by \cite{Tanaka2004}, using first-order, linear calculations which apply to low-mass planets only. For giant planets, it appears more appropriate to use another recipe for eccentricity damping, such as the one provided in Paper I, based on hydrodynamical simulations.      

        In hydrodynamical simulations (included in Paper I), planetary eccentricity growth has sometimes been observed. This is a result of the disc becoming eccentric, and is a long-term effect. The simulations of \citet{Kley2006} stated that a 3-Jupiter mass planet can excite the eccentricity of the disc. Reciprocally, planetary eccentricities could be excited by the disc in some circumstances \citep{Papaloizou2001, Goldreich2003, Ogilvie2003}. Simulations by \citet{Dangelo2006} investigated the long-term evolution of Jupiter-like planets in protoplanetary discs and found eccentricity growth for planets even smaller than 3-Jupiter masses on timescales of several thousand orbits. The growth of the eccentricity happens when the 2:4, 3:5, or 4:6 outer Lindblad resonance becomes dominant \citep{Teyssandier07032016,Duffell2015}. This process therefore depends on the depth and width of the gap in the disc caused by the planet, hence on the planet mass and disc parameters \citep[e.g.,][]{Crida2006}. \citet{Duffell2015} computed a gap opening parameter $\mathcal{K}=q^2/(h^5\alpha)$ from works of different authors, including Paper~I, and showed that in all these studies, eccentricity damping is similarly observed for $\mathcal{K}< 10^{3}$ and eccentricity growth for $\mathcal{K}> 10^{4}$ (their Fig.~9).

%^^^^^^^^^^ Figure 3 ^^^^^^^^^^ %
%^^^^^^^^^^^^^^^^^^^^^^^^^^^^^^ %
 \begin{figure}%[t!]
       \centering
       \includegraphics[width=0.85\hsize]{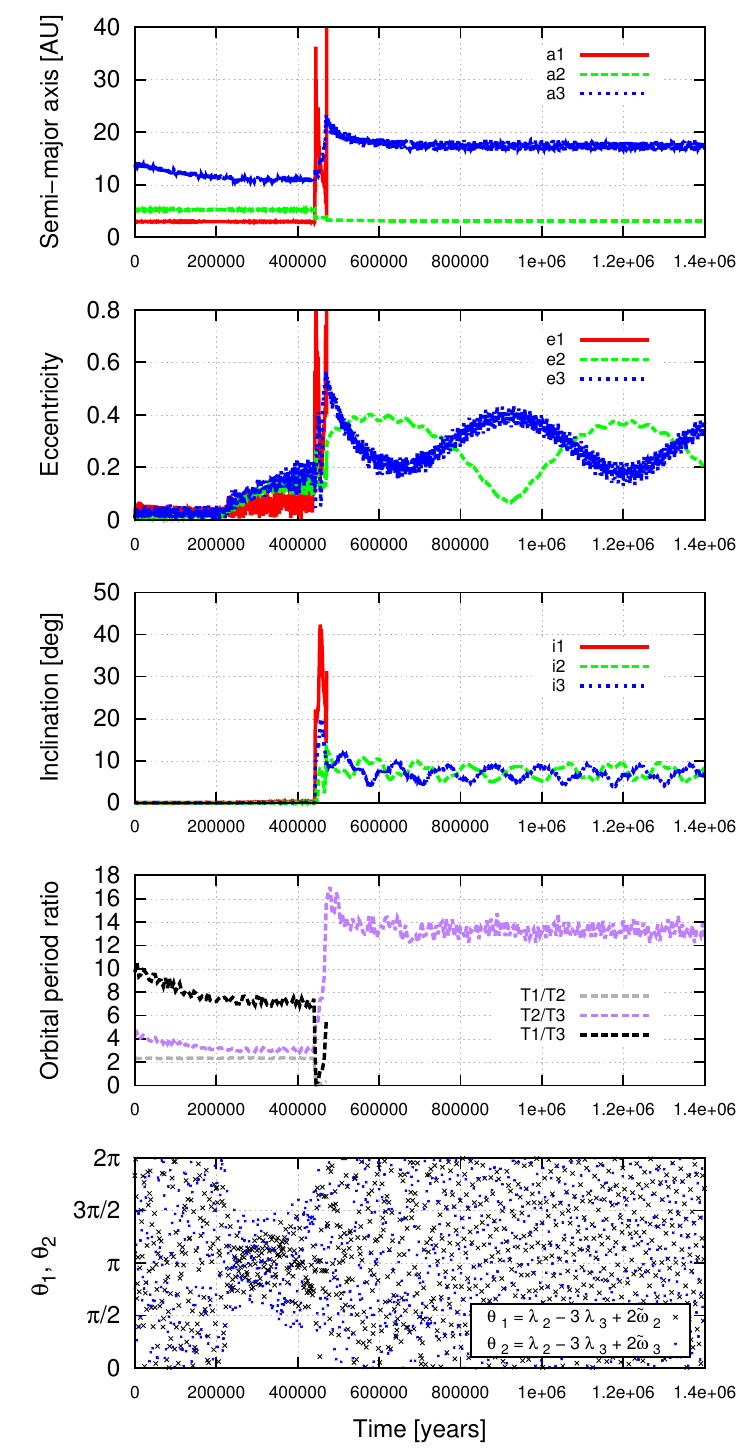} % system index: 3020_157 / M_disc = 16 M_Jup %
       \caption{Inclination excitation due to planet-planet scattering (DM model). After the hyperbolic ejection of the inner planet, the system consists of two planets with a large orbital separation, large eccentricities and slightly inclined orbits. The planetary masses are $m_1 = 1.98$, $m_2 = 8.35$, $m_3 = 3.8$~$M_{\rm Jup}$.
       }
       \label{fig:fig3}
 \end{figure}
%^^^^^^^^^^^^^^^^^^^^^^^^^^^^^^ %
%^^^^^^^^^^^^^^^^^^^^^^^^^^^^^^ %

        The formulae for eccentricity and inclination damping derived in Paper I and used in the present work depend on the eccentricity, inclination, and mass of the planet, as well as on the migration rate via the surface density profile $\Sigma$ (local mass disc $M_{\rm disc}$ in Eqs. \ref{eq_e} and \ref{eq_i}). Previous works using the $K$-prescription argued that the final system configurations are strongly related to the damping chosen (e.g., \citealt{Thommes2003, Libert2009, Libert2011b, Teyssandier2014}). Instead of the $K$-factor dependency, the key parameter of the formulae used in this work is $M_{\rm Disc}$. To check the robustness of our results, we consider four different disc masses in our simulations, being $4, 8, 16,$ and $32 M_{\rm Jup}$, or equivalently four different migration timescales (see Eq. \ref{tII}).

        For consistency, let us give an evaluation of the (initial) ratio of the migration timescale to the eccentricity damping timescale ($\tau_{II}/\tau_{ecc}$) observed in our simulations. The eccentricity damping is in the order of $10^{4} - 10^{5}$ years (depending on the mass of the planet, see Fig. 4 of Paper I), while Type-II migration timescale is of the order of $10^5$ years (see Eq. \ref{tII}), which would result in a $K$ value of $1-10$. This is smaller than the $K$ value assumed in \citet{Lee_Peale}, but in good agreement with measures by \citet{autocrida}. Note that a preliminary study, with a similar three-planet initial set-up as in the present work but making use of the $K$-prescription for the eccentricity damping (no inclination damping considered), was achieved in \cite{Libert2011b} for this range of $K$ values.
Inclination damping timescale $\tau_{inc}$ is in the same range as $\tau_{ecc}$, depending again on the mass of the planet. So the ratio $\tau_{II}/\tau_{inc}$ is in the range 1-10 throughout the evolution of the system.

        Furthermore, let us note that while the damping on the inclinations is either not considered or chosen in an arbitrary way in previous n-body studies, the formulae for eccentricity and inclination damping used in this work have been derived consistently. In particular, it means that no additional free parameter has to be introduced for the inclination damping (as it would be the case for the K-prescription where two K-factors should be considered). For this reason, the formulae of Paper I are appropriate for a study of the impact of the inclination damping on the formation of non-coplanar systems. In the following sections, we identify the dynamical mechanisms producing inclination increase and describe the eccentricity and inclination distributions of planetary systems obtained when adopting these damping formulae.

%^^^^^^^^^^^^^^^^^^^^^^^^^^^^^^ %
%^^^^^^^^^^ Figure 4 ^^^^^^^^^^ %
%^^^^^^^^^^^^^^^^^^^^^^^^^^^^^^ %
 \begin{figure}%[t!]
       \centering
       \includegraphics[width=0.85\hsize]{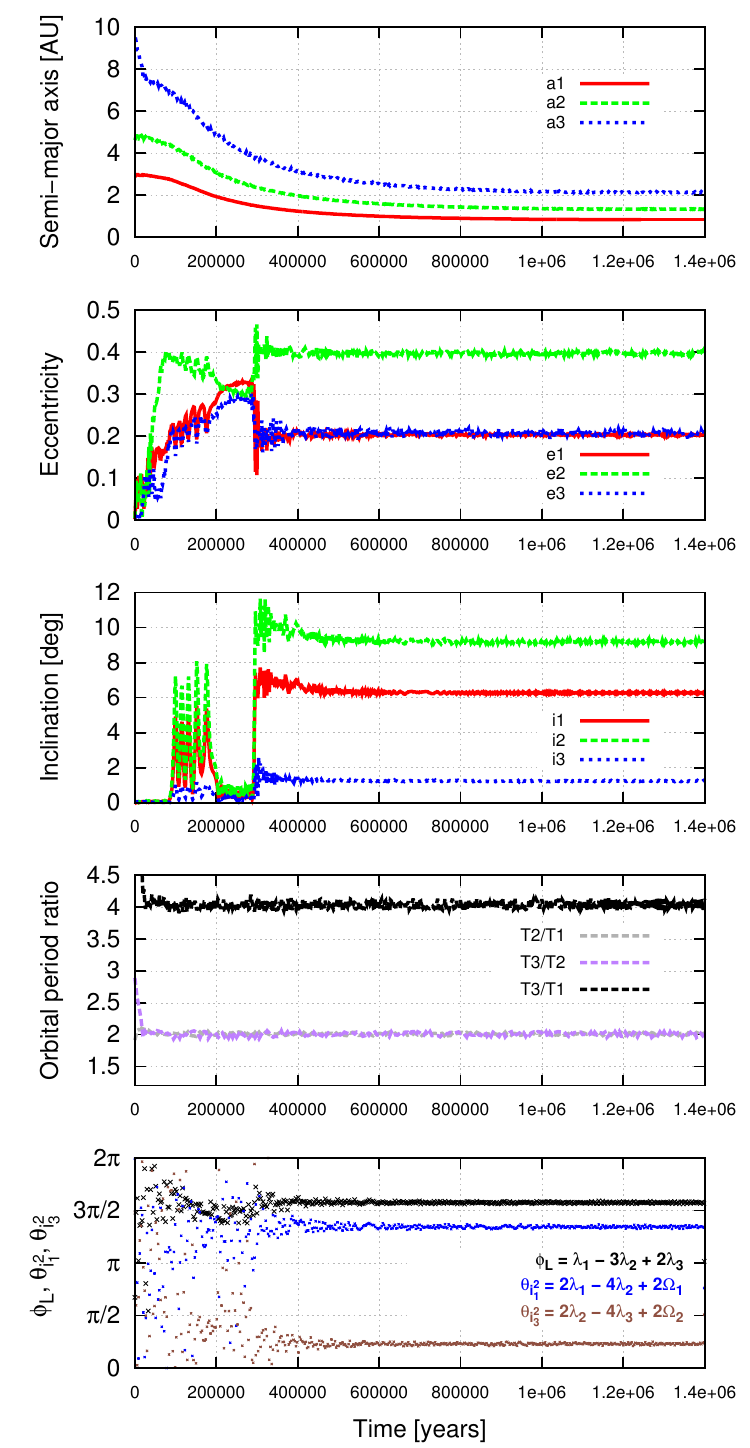} % system index: 3020_157 / M_disc = 16 M_Jup %
       \caption{Inclination excitation due to inclination-type resonance in a Laplace 1:2:4 mean-motion resonance (DM model). The planetary masses are $m_{1}=2.66$, $m_{2}=1.44$ and $m_{3}=0.73$~$M_{\rm Jup}$.}
       \label{fig:fig4}
 \end{figure}
%^^^^^^^^^^^^^^^^^^^^^^^^^^^^^^ %
%^^^^^^^^^^^^^^^^^^^^^^^^^^^^^^ %
%______________________________________________________________________________________________________________
%%%%%%%%%%%%%%%%%%%%%%%%%%%%%%%%%%%%%%%%%%%%%%%%%%%%%%%%%%%%%%%%%%%%%%%%%%%%%%%%%%%%%%%%%%%%%%%%%%%%%%%%%%%%%%%
%%%%%%%%%%%%%%%%%%%%%%%%%%%%%%%%%%%%%%%%%%%%%%%%%%%%%%%%%%%%%%%%%%%%%%%%%%%%%%%%%%%%%%%%%%%%%%%%%%%%%%%%%%%%%%%
%==============================================================================================================

\section{Typical dynamical evolutions}\label{section3}
%==============================================================================================================
%%%%%%%%%%%%%%%%%%%%%%%%%%%%%%%%%%%%%%%%%%%%%%%%%%%%%%%%%%%%%%%%%%%%%%%%%%%%%%%%%%%%%%%%%%%%%%%%%%%%%%%%%%%%%%%
                                %%%%%%%%%%%%%%%%%%%%%%%% Typical evolution %%%%%%%%%%%%%%%%%%%%%%%%
%%%%%%%%%%%%%%%%%%%%%%%%%%%%%%%%%%%%%%%%%%%%%%%%%%%%%%%%%%%%%%%%%%%%%%%%%%%%%%%%%%%%%%%%%%%%%%%%%%%%%%%%%%%%%%% 
        This section shows two typical evolutions of our simulations in the DM model, characterized by inclination excitation. They illustrate well the constant competition between the planet-planet interactions that excite the eccentricity and inclination of giant planets on initially circular and coplanar orbits, and the planet-disc interactions that damp the same quantities. Two dynamical mechanisms producing inclination increase during the disc phase are discussed in the following: planet-planet scattering and inclination-type resonance. The frequency of these dynamical mechanisms will be discussed in Section \ref{4.6}.
        
        The planet-planet scattering mechanism for planetary masses of $m_1 = 1.98$, $m_2 = 8.35$ and $m_3 = 3.8$~$M_{\rm Jup}$ is illustrated in Fig.~\ref{fig:fig3}. While the outer planet migrates, $m_2$ and $m_3$ are captured in a 1:3 mean-motion resonance, at approximately $2\times10^5$~yr, as indicated by the libration of both resonant angles $\theta_1=\lambda_2-3\lambda_3+2\omega_2$ and $\theta_2=\lambda_2-3\lambda_3+2\omega_3$, with $\lambda$ being the mean longitude and $\omega$ the argument of the pericenter. Following the resonance capture, the eccentricities are excited and the two planets migrate as a pair. When $m_2$ approaches a 3:7 commensurability with the inner planet, the whole system is destabilized, leading to planet-planet scattering. Indeed, heavy planets usually lead to dynamical instability before the establishment of a three-body mean-motion resonance \citep{Libert2011b}. The inner less massive body is then ejected from the system at $4\times10^5$ yr, leaving the remaining planets in (slightly) inclined orbits with large eccentricity variations and large orbital separation. As shown here, an excitation in inclinations can result from the planet-planet scattering phase. Let us note here that only during the brief chaotic phase before the ejection of the inner planet, did all the bodies share a common gap, and a reduced rate for eccentricity and inclination damping was applied (see Section 2.2).

%^^^^^^^^^^^^^^^^^^^^^^^^^^^^^^ %
%^^^^^^^^^^ Table 3 ^^^^^^^^^^^ %
%^^^^^^^^^^^^^^^^^^^^^^^^^^^^^^ %
\begin{table} % Discard events table
\begin{center}

  \caption{Percentages of discard events for both disc models at the dispersal of the disc.} \label{tab:title3} 
  \begin{tabular}{ l  l  l  l  l  l }

    \hline
    Model  &  Ejection  & Merging       & Accretion on star     \\ 
    CM     &   43\%       & 48\%             & 9\%              \\ 
    DM     &   61\%       & 38\%             & 1\%              \\ \hline
    
  \end{tabular}
  
\end{center}
\end{table}
%^^^^^^^^^^^^^^^^^^^^^^^^^^^^^^ %
%^^^^^^^^^^^^^^^^^^^^^^^^^^^^^^ %       

%^^^^^^^^^^^^^^^^^^^^^^^^^^^^^^ %       
%^^^^^^^^^^ Table 4 ^^^^^^^^^^^ %
%^^^^^^^^^^^^^^^^^^^^^^^^^^^^^^ %

\begin{table} % Total nbod at the end of the disc
%\hspace{-2cm}
\begin{center}
  \caption{Number of planets in the systems of our simulations at the end of the integration time, for both disc models. The columns show the percentage of systems with $0$, $1$, $2,$ and $3$ planets, respectively. 3\%\ of the simulations are excluded in each modelization, because the final systems consist of (at least) one overly massive planet ($\gtrsim 10 M_{\rm Jup}$, see Section~\ref{section2}).} \label{tab:title4} 
  \begin{tabular}{ l  r  r  r  r  r }

    \hline
    Model                                   & 0 planet                   & 1 planet                      & 2 planets                     & 3 planets                         \\ %\hline
    CM                      & 1\%                        & 25\%                 & 47\%                    & 24\%                          \\ %\hline
    DM - $1.4 \times 10^6$ yr            & 0\%                   & 7\%                  & 50\%                    & 40\%   \\     
    DM - $1.0 \times 10^8$ yr             & 0\%                          & 7\%                   & 50\%                  & 40\%   \\ \hline
    
  \end{tabular}
  
\end{center}
\end{table}
%^^^^^^^^^^^^^^^^^^^^^^^^^^^^^^ %
%^^^^^^^^^^^^^^^^^^^^^^^^^^^^^^ %
          
        Fig. \ref{fig:fig4} shows an example of the establishment of a three-body resonance and the subsequent inclination excitation. In this example, the three planetary masses are $m_{1}=2.66~M_{\rm Jup}$, $m_{2}=1.44~ M_{\rm Jup}$ and $m_{3}=0.73~M_{\rm Jup}$. The two inner planets  are initially in a 1:2 mean-motion resonance. The convergent migration of the outer planet leads to the establishment of a three-body mean-motion resonance, the Laplace 1:2:4 resonance (since $n_2$:$n_1=1$:$2$ and $n_3$:$n_2=1$:$2$, the multiple-planet resonance is labeled as $n_3$:$n_2$:$n_1 = 1$:$2$:$4$), at approximately $2\times 10^4$ yr. The libration of the resonant angle $\phi_L=\lambda_1-3\lambda_2+2\lambda_3$ is shown in the bottom panel of Fig. \ref{fig:fig4}. From then on, the three planets migrate together while in resonance (they never share a common gap) and their eccentricities rapidly increase (up to $0.4$ for $e_2$), despite the damping exerted by the disc. Let us recall that there is no eccentricity excitation from the disc for low-mass planets. When the eccentricities are high enough, the system enters an inclination-type resonance, at approximately $1\times10^5$~yr: the angles $\theta_{i_1^2}=2\lambda_1-4\lambda_2+2\Omega_1$ and $\theta_{i_3^2}=2\lambda_2-4\lambda_3+2\Omega_2$ (where $\Omega$ is the longitude of the ascending node) start to librate and a rapid growth of the inclinations is observed. However, the strong damping exerted by the disc on the planets leads the planets back to the midplane and the system is temporarily outside the inclination-type resonance. On the other hand, the eccentricities of the planets keep increasing as migration continues, in such a way that the system re-enters the inclination-type resonance at $\sim$ $3\times10^5$ yr, with both critical angles $\theta_{\text{i}_1^2}$ and $\theta_{\text{i}_3^2}$ in libration and a new sudden increase in inclination. This time the inclination values is maintained for a long time, since the damping exerted by the gas disc is weaker due to the exponential decay of the disc mass. On the contrary, the weak inclination damping contributes in an appropriate way to the long-term stability of the planetary system by keeping the eccentricity and inclination values constant. This example shows that resonant dynamical interactions between the planets during the disc phase can also form non-coplanar planetary systems. However, when the capture in the inclination-type resonance happens very rapidly, the gas disc is so massive (disc-dominated case) that it forces the planet to get back to the midplane in a very short timescale ($\sim 5\times10^4$ yr). This is the case for the large majority of our simulations (see Section~\ref{section5}).     
%______________________________________________________________________________________________________________
%%%%%%%%%%%%%%%%%%%%%%%%%%%%%%%%%%%%%%%%%%%%%%%%%%%%%%%%%%%%%%%%%%%%%%%%%%%%%%%%%%%%%%%%%%%%%%%%%%%%%%%%%%%%%%%
%%%%%%%%%%%%%%%%%%%%%%%%%%%%%%%%%%%%%%%%%%%%%%%%%%%%%%%%%%%%%%%%%%%%%%%%%%%%%%%%%%%%%%%%%%%%%%%%%%%%%%%%%%%%%%%
%==============================================================================================================

%^^^^^^^^^^^^^^^^^^^^^^^^^^^^^^ %
%^^^^^^^^^^ Figure 5 ^^^^^^^^^^ %
%^^^^^^^^^^^^^^^^^^^^^^^^^^^^^^ %
 \begin{figure}%[t!]
       \centering
        \includegraphics[width=\hsize]{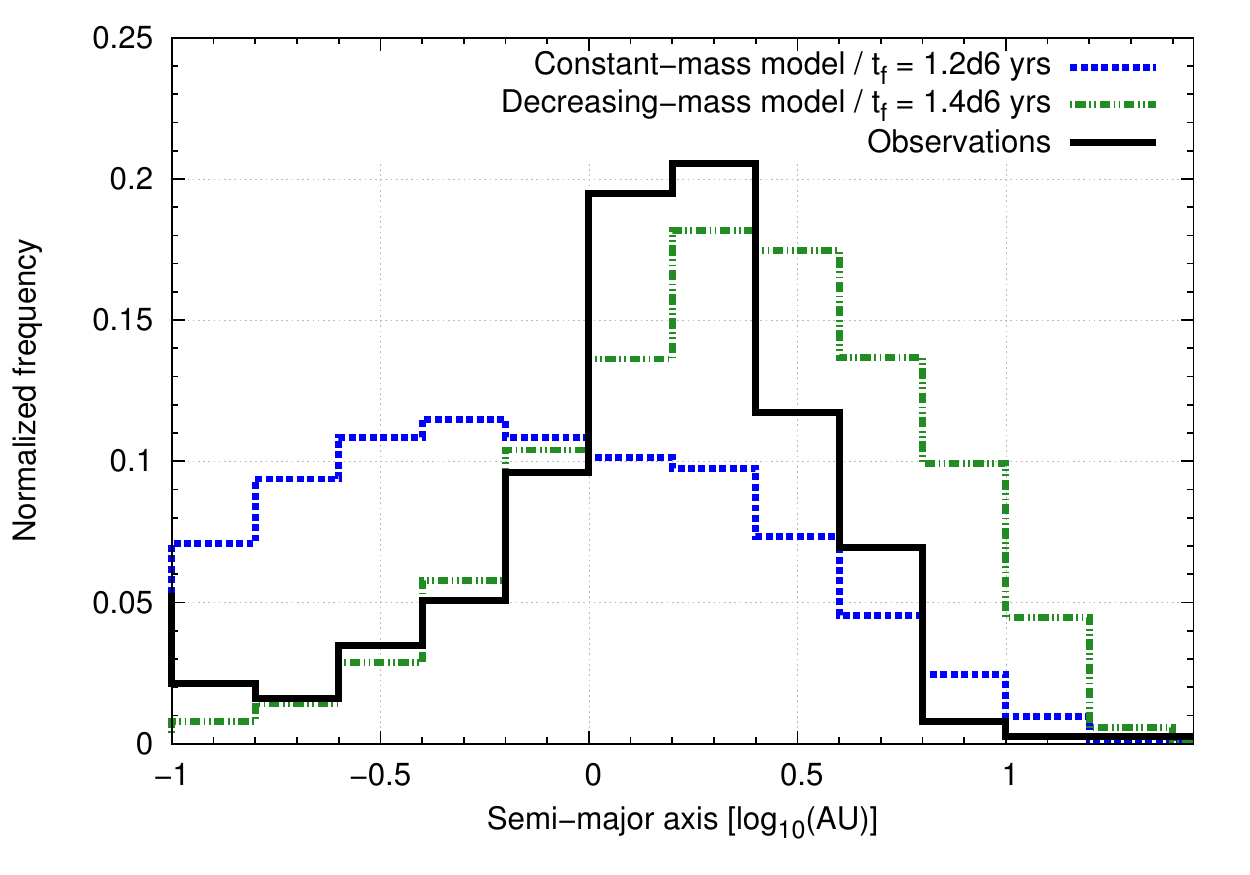}  % Semi-major axis distribution for both disc modelization and 1.4d6 yrs
        \caption{Normalized semi-major axis distributions in both disc modelizations for all the initial disc masses (blue and green lines), and in the observed giant planet population (with $a>0.1$ AU and $M_p~\in~[0.65, 10] M_{\rm Jup}$, black line). The bin size is $\Delta~log(a) = 0.2$.}
       \label{fig:fig5}
 \end{figure}
%^^^^^^^^^^^^^^^^^^^^^^^^^^^^^^ %
%^^^^^^^^^^^^^^^^^^^^^^^^^^^^^^ %

%^^^^^^^^^^^^^^^^^^^^^^^^^^^^^^ %
%^^^^^^^^^^ Figure 6 ^^^^^^^^^^ %
%^^^^^^^^^^^^^^^^^^^^^^^^^^^^^^ %
 \begin{figure}%[t!]
       \centering
       \includegraphics[width=\hsize]{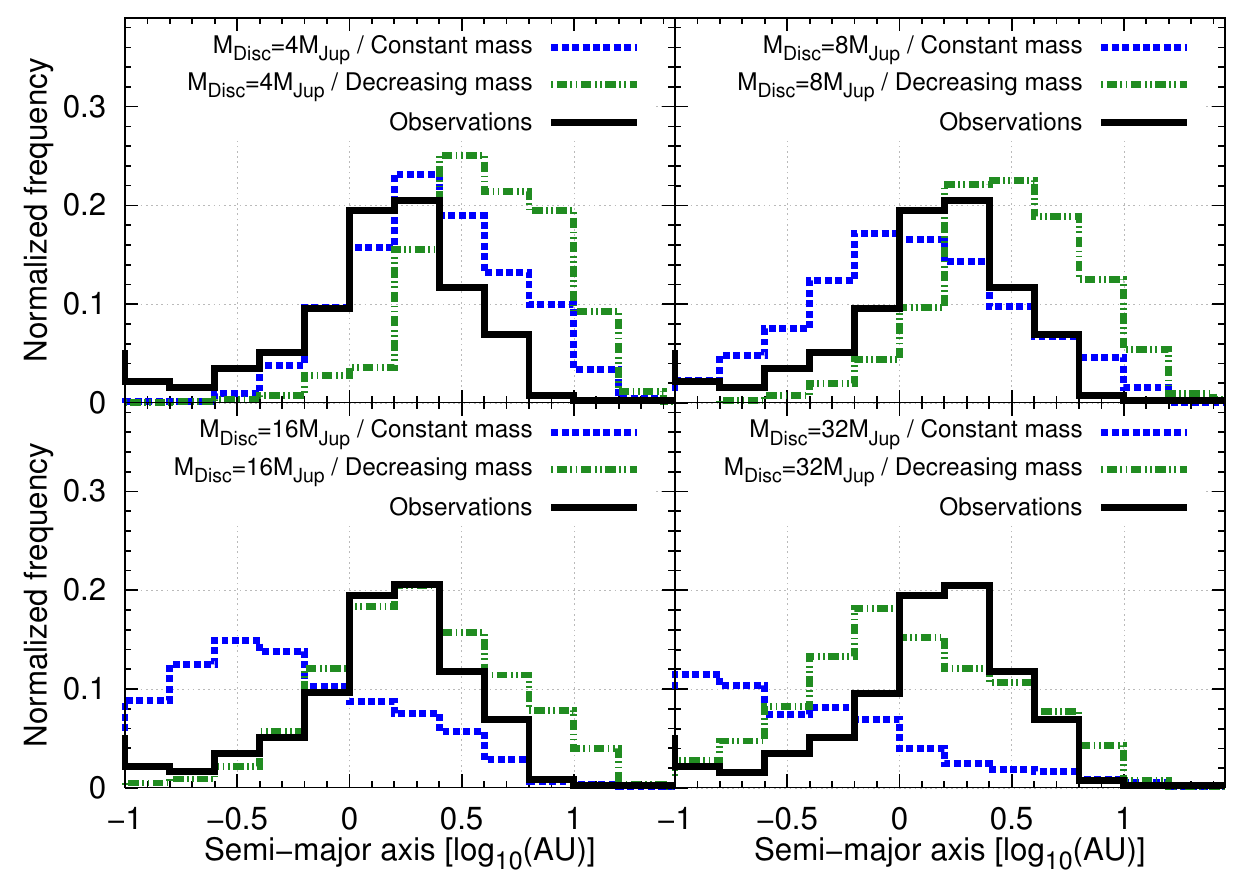}  % Multi-figure for semi-major axis distribution
       \caption{Same as Fig. \ref{fig:fig5} for the different initial disc masses. }
       \label{fig:fig6}
 \end{figure}
%^^^^^^^^^^^^^^^^^^^^^^^^^^^^^^ %
%^^^^^^^^^^^^^^^^^^^^^^^^^^^^^^ %
%______________________________________________________________________________________________________________
%%%%%%%%%%%%%%%%%%%%%%%%%%%%%%%%%%%%%%%%%%%%%%%%%%%%%%%%%%%%%%%%%%%%%%%%%%%%%%%%%%%%%%%%%%%%%%%%%%%%%%%%%%%%%%%
%%%%%%%%%%%%%%%%%%%%%%%%%%%%%%%%%%%%%%%%%%%%%%%%%%%%%%%%%%%%%%%%%%%%%%%%%%%%%%%%%%%%%%%%%%%%%%%%%%%%%%%%%%%%%%%
%==============================================================================================================

\section{Parameter distributions and resonance configurations at the dispersal of the disc}\label{section4}
%==============================================================================================================
%%%%%%%%%%%%%%%%%%%%%%%%%%%%%%%%%%%%%%%%%%%%%%%%%%%%%%%%%%%%%%%%%%%%%%%%%%%%%%%%%%%%%%%%%%%%%%%%%%%%%%%%%%%%%%%
                        %%%%%%%%%%%%%%%%%%%%%%%% Population statistics %%%%%%%%%%%%%%%%%%%%%%%%
%%%%%%%%%%%%%%%%%%%%%%%%%%%%%%%%%%%%%%%%%%%%%%%%%%%%%%%%%%%%%%%%%%%%%%%%%%%%%%%%%%%%%%%%%%%%%%%%%%%%%%%%%%%%%%%
        
        This section describes the main characteristics of the systems formed by our scenario combining migration in the disc and planet-planet interactions. On the one hand, we aim to see whether or not the parameters of the systems formed in our simulations are consistent with the observations, showing how important the role of planet-planet interactions during the disc phase is in the formation of planetary systems. On the other hand, our objective is to study the impact of the inclination damping on the final configurations (Sections \ref{4.1} and~\ref{4.2}), especially on the formation of non-coplanar systems and the resonance captures (Sections \ref{4.3}-\ref{4.5}).

      The systems are followed on a timescale slightly longer than the disc's lifetime, that is $1.2\times10^6$~yr for the $4000$ simulations of the CM model and $1.4\times10^6$~yr for $5800$ simulations of the DM model (see Table~\ref{tab:title1}). This timescale is appropriate for the study of the influence of the disc-planet and planet-planet interactions during the disc phase. The systems emerging from the disc phase would most probably experience orbital rearrangement in the future due to planet-planet interactions. The issue of the long-term evolution and stability of the systems will be addressed in Section \ref{section5}.         

        Let us begin with an overview of our simulations. As the eccentricities of the planets increase due to resonance capture, most of the systems evolve towards close encounters between the bodies, leading to ejections, mergings and/or accretions on the star. The percentages of these three possible outcomes are given in Table \ref{tab:title3}, for both disc models. In the DM model, ejections are preferred over collisions among the planets and with the central star. 

        Table \ref{tab:title4} shows the final number of planets at the end of the integration time. The constant-mass model preferably forms systems with a small number of planets, since migration is efficient during all the simulations (disc-dominated case favored) and pushes the planets closer to one another and closer to the parent star. However, the more realistic decreasing-mass scenario shows that systems with  two or three planets are the most usual outcomes: $50\%$ of our systems consist of two planets at the end of the integration time and $40\%$ of the systems are still composed of three planets.
        
        Hereafter, we describe the orbital characteristics of the planetary systems formed by our mechanism. Semi-major axes, eccentricities, and inclinations are discussed in Sections \ref{4.1}, \ref{4.2}, and \ref{4.3}, respectively. Section \ref{4.4} analyzes the mutual Hill separation of the orbits, while three-body resonances are studied in Section \ref{4.5}. Unless otherwise stated, our analysis relates to the planetary systems of the DM model with $a>0.1$ AU (because tidal/relativistic effects are not included in our work) and mass in the interval $[0.65,10]M_{\rm Jup}$, as they emerge from the disc phase (at $1.4\times10^6$ yr), namely $13036$~planets in $5644$~systems. Comparisons are made with the exoplanets of the observational data that suffer from the same limitations in semi-major axis and mass.

%%%%%%%%%%%%%%%%%%%%%   Semi-major axis  %%%%%%%%%%%%%%%%%%%%%
\subsection{Semi-major axes}\label{4.1}
        We first investigate the effect of the gas disc on the final distribution of semi-major axis. Fig. \ref{fig:fig5} shows the normalized semi-major axis distribution (in logarithmic scale), for all the systems of the DM model as they emerge from the gas phase (green dashed line). For completeness, the CM model is also added to the plot (blue dashed line) in order to evaluate the efficiency of the orbital migration. Indeed, the efficiency of the migration mechanism depends on the amount of surrounding gas in the vicinity of the planet, and the planets end up with smaller semi-major axis in the CM modelization, as expected. The black solid line represents the observed giant planet population (we have excluded the planets with $a<0.1$ AU and mass out of the interval $[0.65,10]M_{\rm Jup}$). Unlike the CM model, the DM semi-major axis distribution and the one of the observations follow a similar trend.  

%^^^^^^^^^^^^^^^^^^^^^^^^^^^^^^ %
%^^^^^^^^^^ Figure 7 ^^^^^^^^^^ %
%^^^^^^^^^^^^^^^^^^^^^^^^^^^^^^ %
 \begin{figure}%[t!]
       \centering
       \includegraphics[width=\hsize]{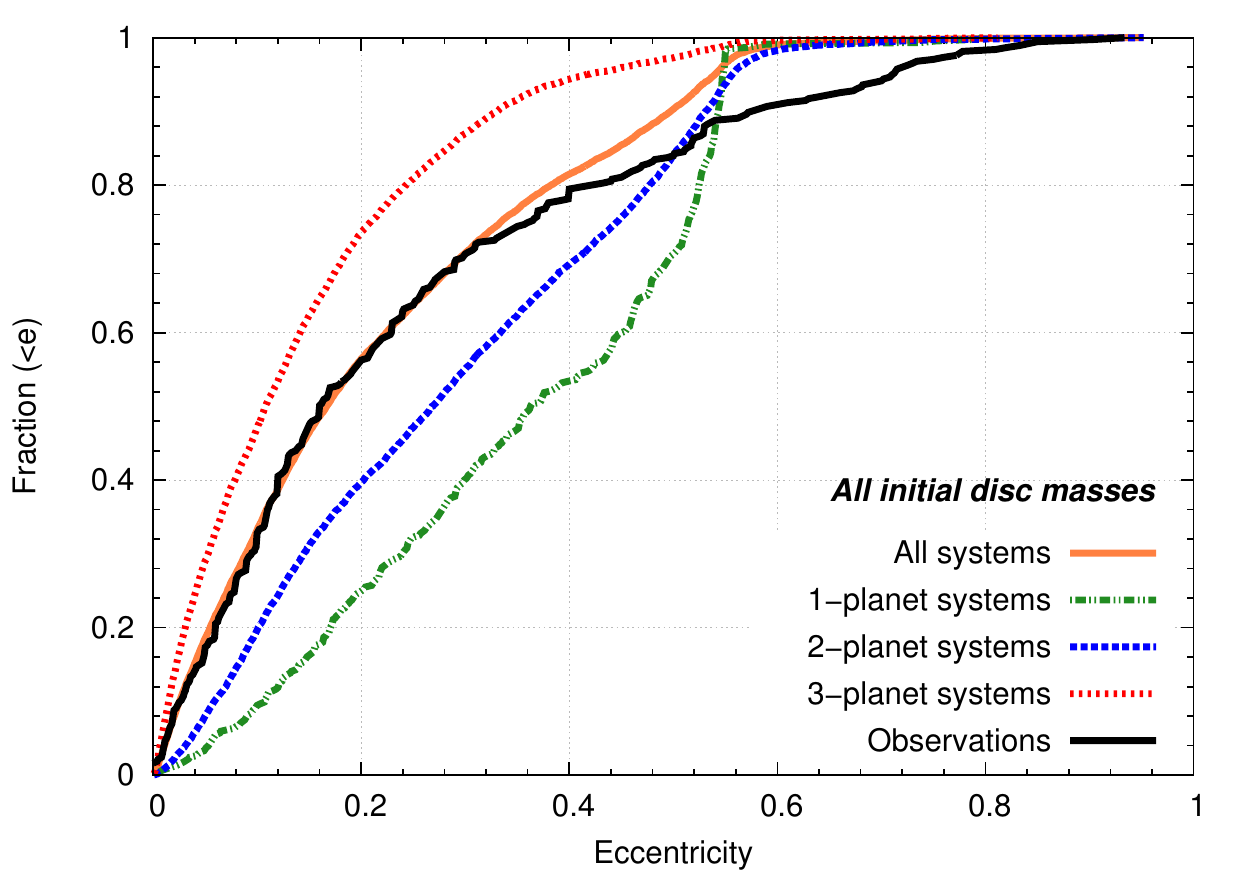}  % Multi-figure for semi-major axis distribution
       \includegraphics[width=\hsize]{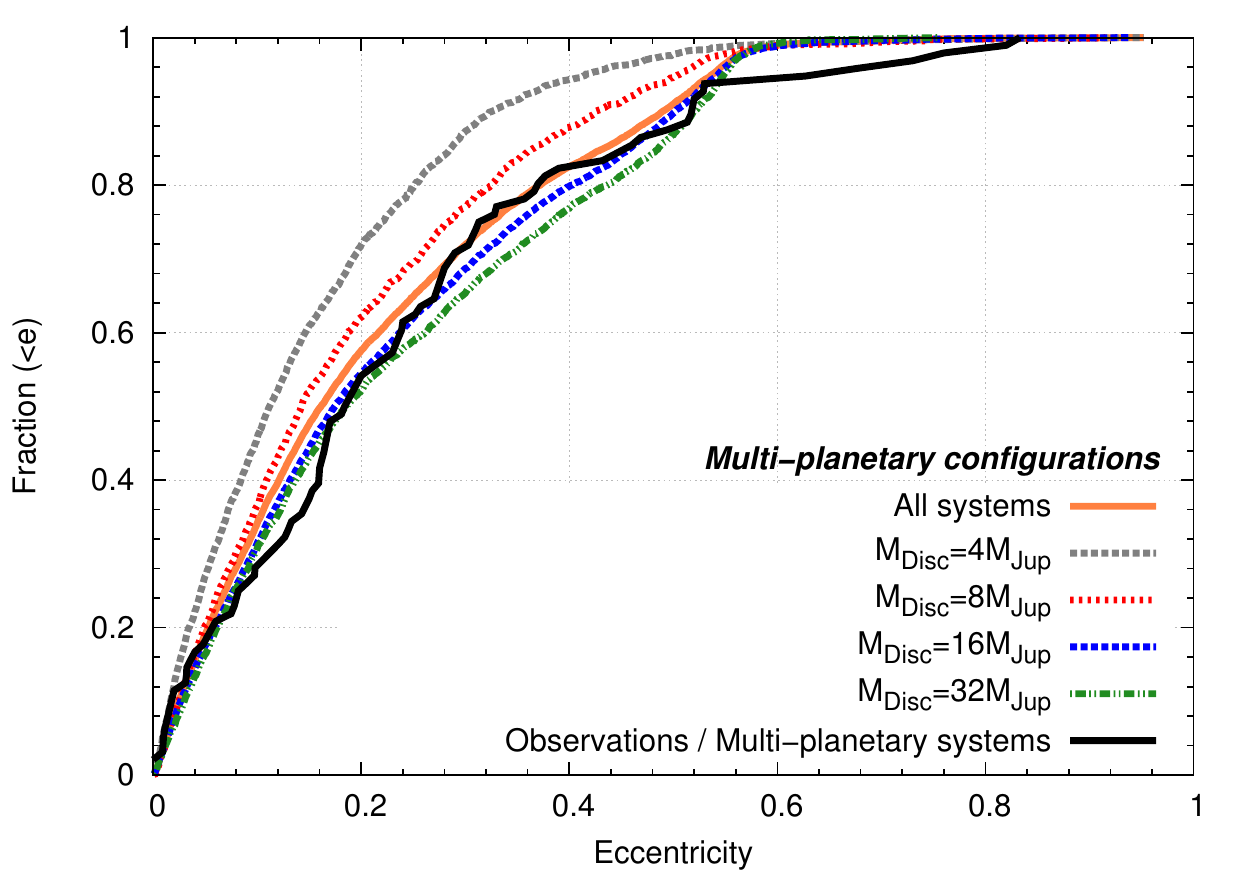}  % Multi-figure for semi-major axis distribution
       \caption{Cumulative eccentricity distributions at the dispersal of the disc. The black lines correspond to the observations. {\it Top panel:} The dependence of the distribution of all systems on the number of planets. {\it Bottom panel:} The dependence of the distribution of multi-planetary systems on the initial disc mass.}
       \label{fig:fig7}
 \end{figure}
%^^^^^^^^^^^^^^^^^^^^^^^^^^^^^^ %
%^^^^^^^^^^^^^^^^^^^^^^^^^^^^^^ %

%^^^^^^^^^^^^^^^^^^^^^^^^^^^^^^ %
%^^^^^^^^^^ Figure 9 ^^^^^^^^^^ %
%^^^^^^^^^^^^^^^^^^^^^^^^^^^^^^ %
 \begin{figure}%[t!]
       \centering
       \includegraphics[width=\hsize]{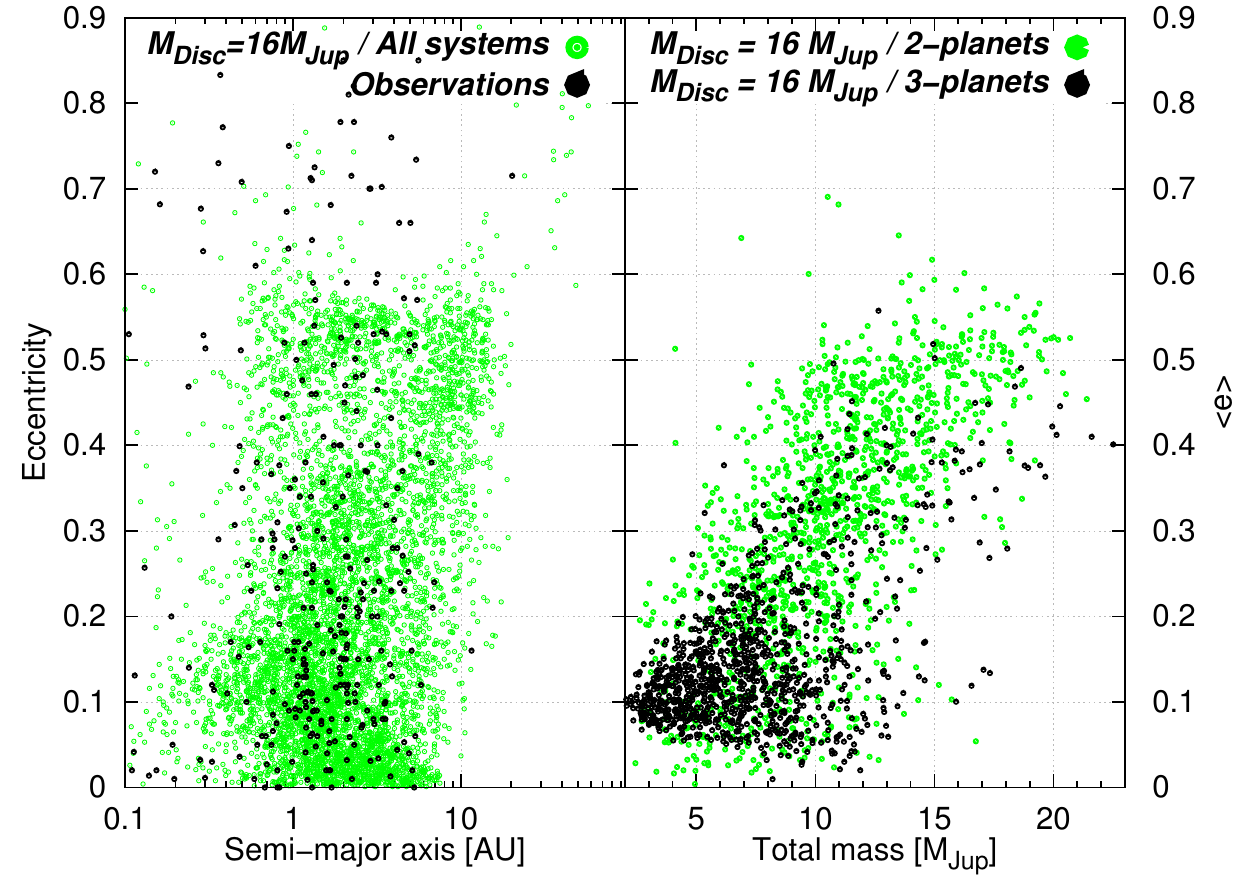}  % Multi-figure for semi-major axis distribution
       \caption{\textit{Left panel}. Semi-major axis vs. eccentricity for our simulations with $M_{\rm Disc}=16M_{\rm Jup}$ (green dots) and the observations (black dots). \textit{Right panel}: Total mass vs. mean eccentricity for each multi-planetary configuration of our $16M_{\rm Jup}$ disc simulations. Black dots correspond to three-planet systems and green dots to two-planet systems.}
       \label{fig:fig9}
 \end{figure}
%^^^^^^^^^^^^^^^^^^^^^^^^^^^^^^ %
%^^^^^^^^^^^^^^^^^^^^^^^^^^^^^^ %

        We further analyze the DM model semi-major axis distribution, investigating the influence of the initial disc mass in Fig.~\ref{fig:fig6}. The different panels present the normalized semi-major axes distribution (in logarithmic scale) for initial disc mass of $4$, $8$, $16,$ and $32$ $M_{\rm Jup}$, respectively. For more massive discs, the migration mechanism is more efficient and the planets evolve closer to the parent star. Let us note that the best agreement between the simulated and observed distributions is obtained for the $16$ $M_{\rm Jup}$ disc (bottom left panel of Fig. \ref{fig:fig6}).   
%%%%%%%%%%%%%%%%%%%%%%%%%%%%%%%%%%%%%%%%%%%%%%%%%%%%%%%%%%%%%%

%%%%%%%%%%%%%%%%%%%%%   Eccentricity  %%%%%%%%%%%%%%%%%%%%%%%%

\subsection{Eccentricities}\label{4.2}
         We find good agreement between the eccentricities of the planetary systems obtained by our simulations and the detected ones, as shown in Fig. \ref{fig:fig7}, where both cumulative eccentricity distributions are displayed (top panel). We see that the eccentricities observed in our simulations are well diversified and present the same general pattern as the observed eccentricities. In particular, the similarity of the curves for eccentricities smaller than $0.35$ is impressive. However, our model underproduces highly eccentric orbits: only $\sim$~$9.7 \%$ of the planets have an eccentricity higher than~$0.5$ in our simulations. We highlight that the modelization of the damping formulae is not accurate for $e>2/3$ (Section \ref{section2}). Also, the orbital elements of our systems are considered immediately after the dispersal of the disc ($1.4\times10^6$ yr) and we show in Section \ref{section5} that orbital adjustments due to planet-planet interactions can still come into play afterwards. 

        Furthermore, it is well known that part of the high eccentricities reported in the observations could originate from additional mechanisms not considered here, such as the gravitational interactions with a distant companion (binary star or massive giant planet). These highly eccentric planets are mostly in single planet systems, at least in the observational data for a distance of $30$ AU from the star. For this reason, in the bottom panel of Fig. \ref{fig:fig7}, we focus  on the eccentricities of (final) multi-planetary systems only. We first observe that removing the single-planet systems reduces the disagreement at high eccentricities, in particular from $0.35$ up to $0.55$, higher eccentricities not being correctly modelized by our study. 
        
        The cumulative eccentricity distributions for the four initial disc masses considered in this study are also shown in the bottom panel of Fig. \ref{fig:fig7} (dashed lines). The lower the mass of the disc, the lower the eccentricities of the multi-planetary configurations observed in the simulations. The initial disc mass of $16$~$M_{\rm Jup}$ gives the best approximation of the observational distribution. To put this in perspective, let us add that the mean value of the total mass of the planetary systems at the beginning of the simulations is $10.5$~$M_{\rm Jup}$. Since our approximation well matches the observations for small and medium eccentricities, this would imply that the observed extrasolar systems are consistent with an initial disc mass comparable to the total mass of the planets initially formed in the disc.     

        The correlation between semi-major axis and eccentricity is examined in the left panel of Fig. \ref{fig:fig9}. In this analysis, we consider the $16M_{\rm Jup}$ disc simulations, since the previous figures suggest that this initial disc mass presents the best curve fittings for both the semi-major axis and the eccentricity independently. The left panel of Fig. \ref{fig:fig9} shows the matching between our simulations and the observations in the semi-major axis vs. eccentricity graph. The results are in agreement with the study of \citet{Matsumura2010}, combining N-body dynamics with hydrodynamical disc evolution.
        
        Finally, the mean eccentricity for each system that ends up in a multi-planetary configuration in our simulations is shown in the right panel of Fig. \ref{fig:fig9} as a function of the total mass of the system. For the systems consisting of two planets at the dispersal of the disc, the higher the total mass of the system, the higher the mean eccentricity. Indeed, the eccentricities are excited by mean-motion resonance capture, planet-planet scattering and interactions with the disc. These mechanisms are more efficient for massive planets. In particular, concerning the last one, the disc, instead of damping the planetary eccentricities, can induce eccentricity excitation \citep{Papaloizou2001,Dangelo2006} for high-mass planets ($M_{p} \gtrsim 5.5 M_{\rm Jup}$). This feature is included in the damping formulae derived in Paper I. For the three-planet systems, the mean eccentricities are smaller, since planet-planet scattering did not take place and only the other two mechanisms operated. 

        We do not discuss here the eccentricities of the CM model as no significant difference was observed. The same holds true for the inclinations. 

%^^^^^^^^^^^^^^^^^^^^^^^^^^^^^^ %
%^^^^^^^^^^ Figure 10 ^^^^^^^^^ %
%^^^^^^^^^^^^^^^^^^^^^^^^^^^^^^ %
 \begin{figure}%[t!]
       \centering
       \includegraphics[width=\hsize]{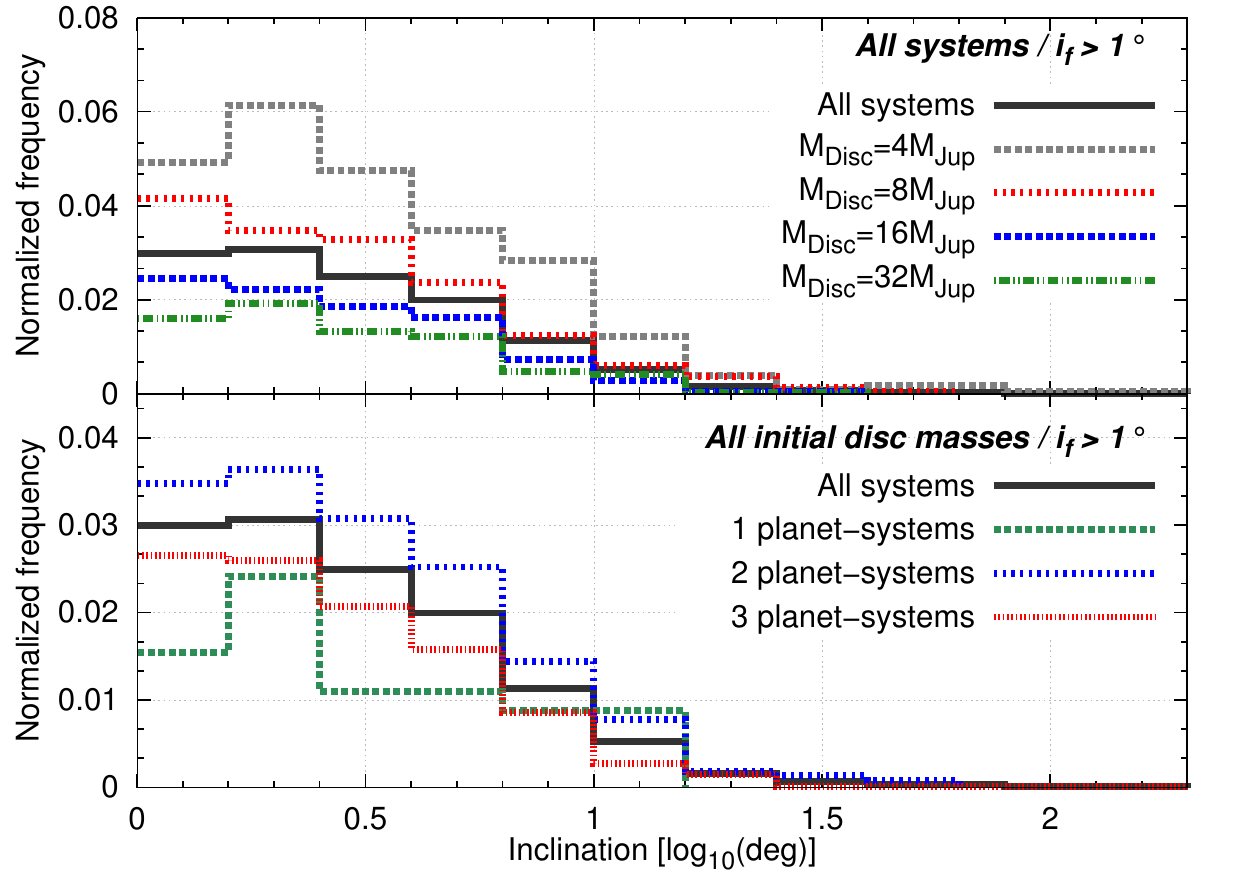}  % Multi-figure for semi-major axis distribution
       \caption{Normalized inclination distribution for all the planets with $i >~1^{\circ}$ at the dispersal of the disc. {\it Top panel:} The dependence of the distribution on the initial disc mass. {\it Bottom panel:} The dependence of the distribution on the number of planets in the (final) systems. The bin size is $\Delta$ log($i$) = 0.2.}
       \label{fig:fig10}
 \end{figure}
%^^^^^^^^^^^^^^^^^^^^^^^^^^^^^^ %
%^^^^^^^^^^^^^^^^^^^^^^^^^^^^^^ %
        
%^^^^^^^^^^^^^^^^^^^^^^^^^^^^^^ %
%^^^^^^^^^^ Figure 12 ^^^^^^^^^ %
%^^^^^^^^^^^^^^^^^^^^^^^^^^^^^^ %
 \begin{figure}%[t!]
       \centering
       \includegraphics[width=\hsize]{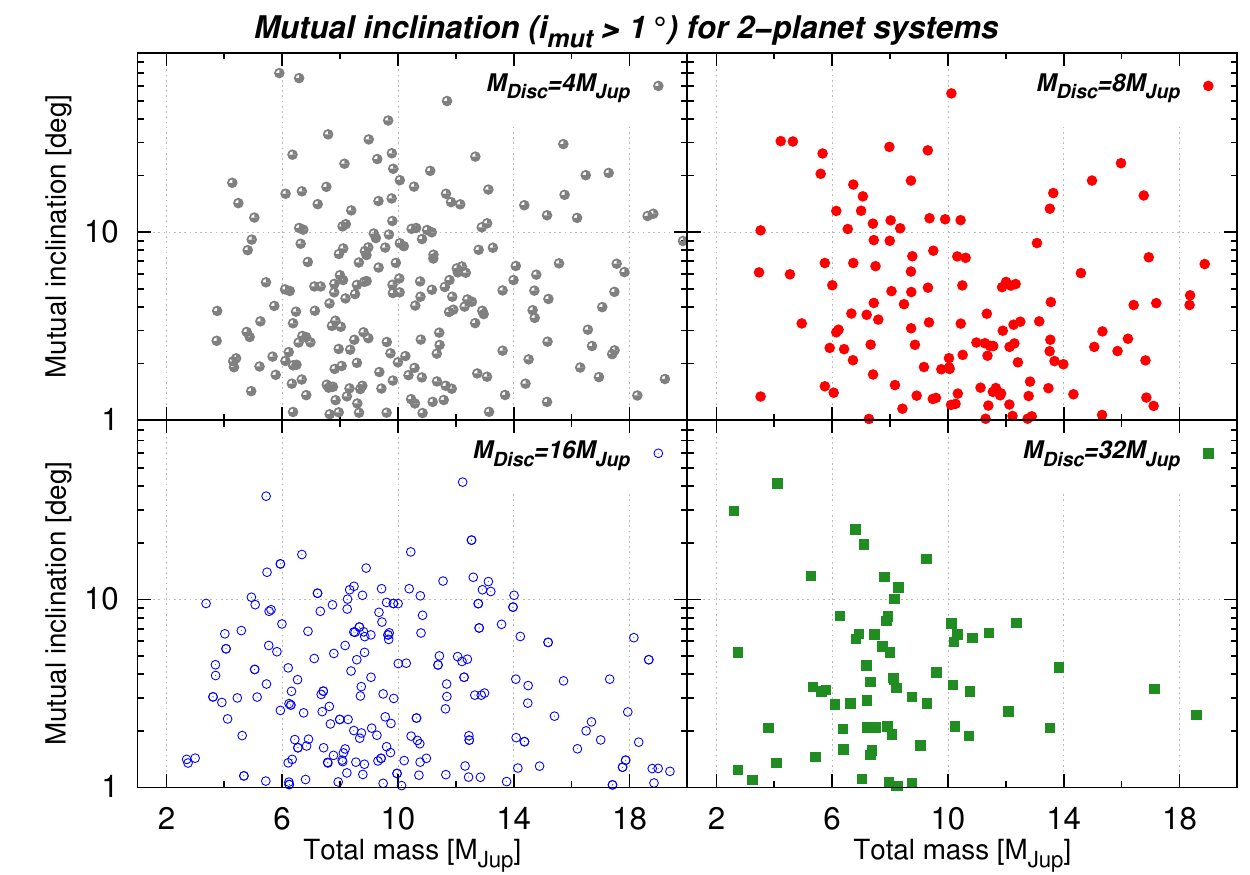}  % Multi-figure for semi-major axis distribution
       \caption{Mutual inclination of the two-planet systems (in logarithmic scale) as a function of the total mass of the planets in the system for the four different disc masses.}
       \label{fig:fig12}
 \end{figure}
%^^^^^^^^^^^^^^^^^^^^^^^^^^^^^^ %
%^^^^^^^^^^^^^^^^^^^^^^^^^^^^^^ %       
%%%%%%%%%%%%%%%%%%%%%%%%%%%%%%%%%%%%%%%%%%%%%%%%%%%%%%%%%%%%%%

%%%%%%%%%%%%%%%%%%%%%   Inclinations  %%%%%%%%%%%%%%%%%%%%%%%%
\subsection{Inclinations}\label{4.3}    
%^^^^^^^^^^^^^^^^^^^^^^^^^^^^^^ %
%^^^^^^^^^^ Figure 13 ^^^^^^^^^ %
%^^^^^^^^^^^^^^^^^^^^^^^^^^^^^^ %
 \begin{figure}%[t!]
       \centering
       \includegraphics[width=\hsize]{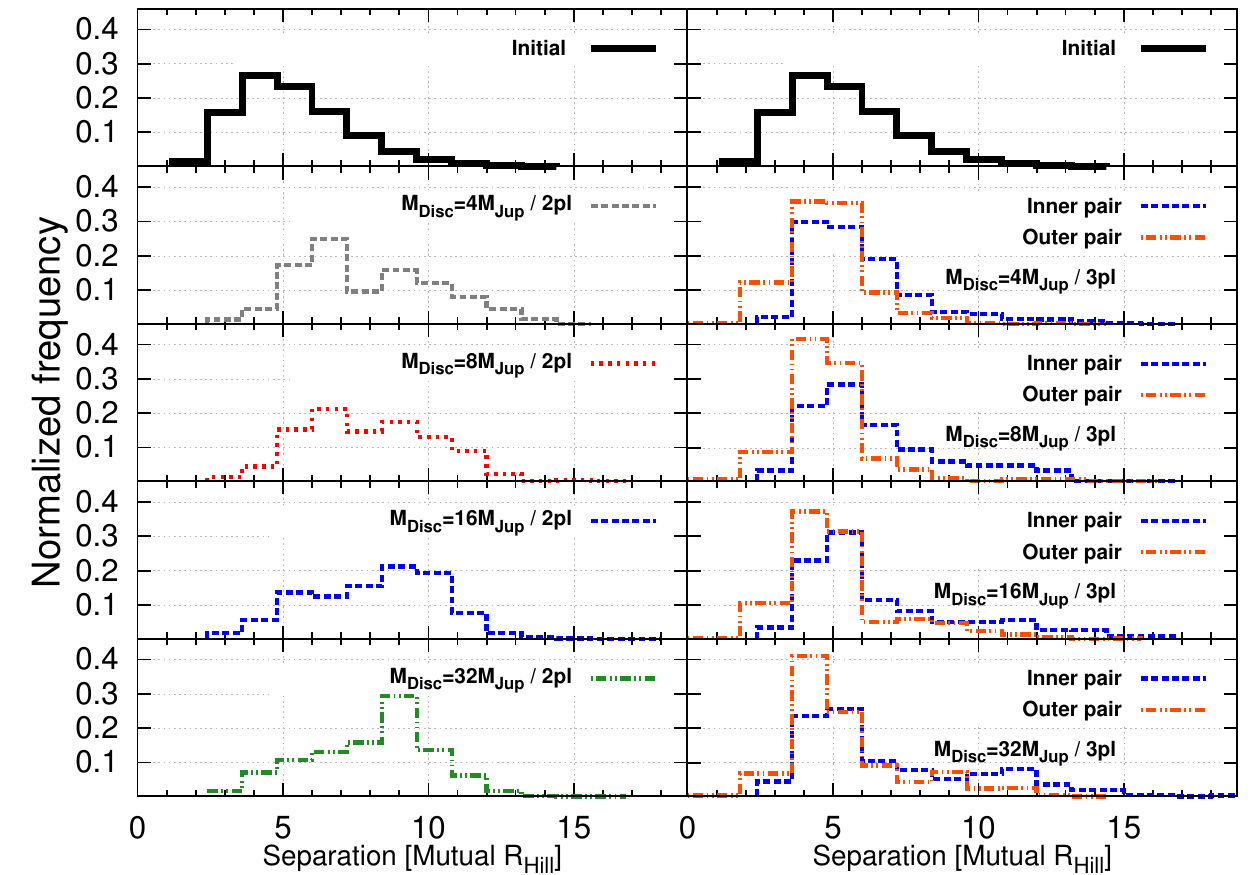}  % Multi-figure for semi-major axis distribution
       \caption{The initial and final Hill neighbor separation normalized distributions ($D_{\rm Hill} = (\Delta a_{i,j}) / R_{\rm Hill,a_{i},a_{j}}$) of the systems of our simulations. The left column displays the two-planet systems and the right column the inner and outer pairs of the three-planet systems. Each panel represents a different initial disc mass. The bin size is $\Delta D_{\rm Hill} = 1.2$.}
       \label{fig:fig13}
 \end{figure}
%^^^^^^^^^^^^^^^^^^^^^^^^^^^^^^ %
%^^^^^^^^^^^^^^^^^^^^^^^^^^^^^^ %

        In Fig. \ref{fig:fig10} we present the normalized inclination distribution (black solid line), as a function of the initial disc mass (top panel), and the number of planets in the (final) systems (bottom panel). Only the planets with $i > 1^{\circ}$ are represented. A clear outcome from the top panel is that the smaller the initial disc mass, the larger the fraction of planets with higher inclinations in the timescale of disc's lifetime. For a $4$~$M_{\rm Jup}$ disc, $\sim 24\%$ of the planets have $i > 1^{\circ}$, while this percentage drops dramatically to approximately $7\%$ for a $32$~$M_{\rm Jup}$ disc. It is simply explained by the fact that a more massive disc exerts a stronger damping on the inclination of the planet. 
        
        The normalized inclination distribution (when $i > 1^{\circ}$) also strongly depends on the final number of bodies in the system after the gas phase, as shown in the bottom panel of Fig. \ref{fig:fig10}. As expected, higher inclinations are typically found for two-planet systems, since these systems have undergone a scattering event during the gas phase. Moreover, regarding the mutual inclination of two-planet systems, there is no correlation between the total mass of the system and the mutual inclination of the planetary orbits. In Fig. \ref{fig:fig12} we show that, for the four disc masses considered here, mutual inclinations can be pumped regardless of the total mass of the systems.        
%%%%%%%%%%%%%%%%%%%%%%%%%%%%%%%%%%%%%%%%%%%%%%%%%%%%%%%%%%%%%%%%%

%%%%%%%%%%%%%%%%%%%%%     Hot jupiters      %%%%%%%%%%%%%%%%%%%%%
\subsection{Hot Jupiters}\label{4.4.0}
        Concerning hot Jupiters, $\sim\!2\%$ of the planets are found with semi-major axes in the range $[0.02,0.2]$ AU after the dispersal of the disc. We remind the reader that our model is not accurate for planets with $a<0.1$ AU, since we do not consider tidal/relativistic effects. However, our simulations seem to indicate that planet-planet interactions during migration in the protoplanetary disc can produce hot Jupiters on eccentric orbits while the migration of single planets would leave them circular. In our results, $\sim\!\!24\%$ of these close-in planets have eccentricities higher than $0.2$. This suggests that planetary migration, including planet-planet interactions, is a valid mechanism to produce both circular and eccentric hot Jupiters.
        
        Moreover, it appears difficult to excite their inclinations larger than $10^\circ$ with respect to the midplane of the disc (only $\sim\!12\%$ of them reach inclination higher than $1^{\circ}$ in our simulations). In this frame, the observed misalignment of approximately half of the hot Jupiters with respect to the stellar equator could be the result of a misalignment between the old disc midplane and the present stellar equator (see for instance \citealt{Batygin}).
%%%%%%%%%%%%%%%%%%%%%%%%%%%%%%%%%%%%%%%%%%%%%%%%%%%%%%%%%%%%%%%%%

%%%%%%%%%%%%%%%%%%%%%   Separations R_Hill  %%%%%%%%%%%%%%%%%%%%%
\subsection{Mutual Hill separation}\label{4.4}
      The \textit{mutual Hill radius} of two planets is defined, by \citet{Gladman1993}, as:
      \begin{equation}
            R_{\rm Hill,a_{i},a_{j}}  =  \left(\frac{m_{i}+m_{j}}{3 M_{\rm star}}\right)^{1/3} \left(\frac{a_{i}+a_{j}}{2}\right),
      \end{equation}
      and we express the {\it Hill neighbor separation} of two planetary orbits as $ D_{\rm Hill} = (a_{j}-a_{i}) / R_{\rm Hill,a_{i},a_{j}} $. In Fig. \ref{fig:fig13}, we show the initial and final Hill neighbor separations of the systems of our simulations. Systems ending up with two-planets are shown in the left column, while the right column shows both inner and outer pairs of the three-planet systems with different colored dashed lines. Each panel corresponds to a different initial value of the mass of the disc. 
      
      It is clear that the three-planet configurations are more compact compared with systems that suffered scattering events. Also, the distribution of the inner pair is slightly wider than the one of the outer pair since only the outer planet is migrating inwards due to the interaction with the disc in our simulations. Concerning the Hill neighbor separation distribution of the two-planet systems, we see that the higher the initial disc mass, the larger the mean separation. 
%%%%%%%%%%%%%%%%%%%%%%%%%%%%%%%%%%%%%%%%%%%%%%%%%%%%%%%%%%%%%%%%%

%^^^^^^^^^^^^^^^^^^^^^^^^^^^^^^ %
%^^^^^^^^^^ Table 5 ^^^^^^^^^^^ % ! from MMR section
%^^^^^^^^^^^^^^^^^^^^^^^^^^^^^^ %
\begin{table} % Decreasing mass simulations table
\begin{center}

\caption{Three-body mean-motion resonances at the dispersal of the disc for both disc modelizations.}
\label{tab:title5} 
  \begin{tabular}{  r  r  r }

    \hline
            Resonance                   &  CM model     & DM model          \\ %\hline
                1:2:4                           &       49\%                    & 54\%    \\ %\hline
                        1:3:6                   &       23\%                    & 19\%                                    \\ %\hline
                        1:2:6                           &       6.5\%                   & 3\%                             \\ %\hline
                        2:5:10                          &       17\%                    & 20\%                                    \\ %\hline
                        2:5:15                          &       1.5\%                   & 1.5\%                                   \\ %\hline
                        other                           &       3\%                     & 2.5\%                                   \\ \hline
                            
  \end{tabular}
\end{center}
\end{table}
%^^^^^^^^^^^^^^^^^^^^^^^^^^^^^^ %
%^^^^^^^^^^^^^^^^^^^^^^^^^^^^^^ %

%%%%%%%%%%%%%%%%%%   Mean-motion reaonances  %%%%%%%%%%%%%%%%%
\subsection{Three-body resonances}\label{4.5}
        As previously mentioned, $40\%$ of the systems (in our DM model simulations) end up in a three-planet configuration (Table \ref{tab:title4}). Depending on the initial separation of the planets, the systems are generally captured in two- or three-body mean-motion resonances during the migration phase. These resonances can either survive until the end of the disc phase or be disrupted in an instability phase. For the systems consisting of three planets at the end of the disc phase, $\sim 10\%$ of them are not in mean-motion resonance, $\sim 25\%$ are in a two-body mean-motion resonance and $\sim 65\%$ are in a three-body mean-motion resonance. This high percentage ($\sim 90\%$) of resonant systems is also observed in \citet{Matsumura2010}. Table \ref{tab:title5} shows that half of the three-body resonant systems are in a 1:2:4 Laplace configuration. The $n_{3}$:$n_{2}$:$n_{1}=1$:$3$:$6$ and $n_{3}$:$n_{2}$:$n_{1}=2$:$5$:$10$ resonances are the second most common configurations. These percentages being the same in both disc modelizations shows that\ unlike the semi-major axis distribution, the establishment of the resonant three-body configurations is not affected by the decrease of the gas disc.      
        
%^^^^^^^^^^^^^^^^^^^^^^^^^^^^^^ %
%^^^^^^^^^^ Table 5 ^^^^^^^^^^^ % ! ATTENTION : from MMR section
%^^^^^^^^^^^^^^^^^^^^^^^^^^^^^^ %
\begin{table} % Decreasing mass simulations table
\begin{center}

\caption{Dynamical history of highly mutually inclined systems. The first column shows all the possible outcomes. For two-planet systems, ejection/collision induced by planet-planet scattering, ejection or collision followed by an inclination-type resonance and ejection/collision after a three-body resonance are the possible scenarios. For three-planet systems, orbital re-arrangement followed or not by a three-body resonance, three-body resonance and orbital re-arrangement after capture in three-body resonance are the four scenarios. The second column gives the number of planets at the end of simulation. The last two columns give the percentages of each scenario for different integration times.}
\label{tab:history} 
  \begin{tabular}{  l r  r  r }

    \hline
            Dynamical history           & \#    &  $1.4 \times 10^6$ yr & $1.0 \times 10^8$ yr    \\ 
            \hline
Ejection/Collision              & 2 &   38\%                    & 45\%  \\ %\hline
Ejection/Collision + MMR        & 2 &   20\%                    & 16\%                                   \\ %\hline
3-B reso + ejection/collision      &2   &       5\%                     & 17\%                                    \\ %\hline
\hline
Orb. instability        &3      &       9\%                             & 8\%                             \\ %\hline
Orb. instability + 3-B reso &3  &       17\%                    & 8\%                                    \\ %\hline
3-B reso        &3      &       10\%                    & 6\%                                    \\ %\hline
3-B reso + orb. instability     & 3 &   1\%                     & 0\%                                    \\ %\hline
                                                    
    \hline
  \end{tabular}
\end{center}
\end{table}
%^^^^^^^^^^^^^^^^^^^^^^^^^^^^^^ %
%^^^^^^^^^^^^^^^^^^^^^^^^^^^^^^ %       

%^^^^^^^^^^^^^^^^^^^^^^^^^^^^^^ %
%^^^^^^^^^^ Figure 14 ^^^^^^^^^ %
%^^^^^^^^^^^^^^^^^^^^^^^^^^^^^^ %
 \begin{figure*}%[t!]
       \centering
       \includegraphics[width=0.24\hsize]{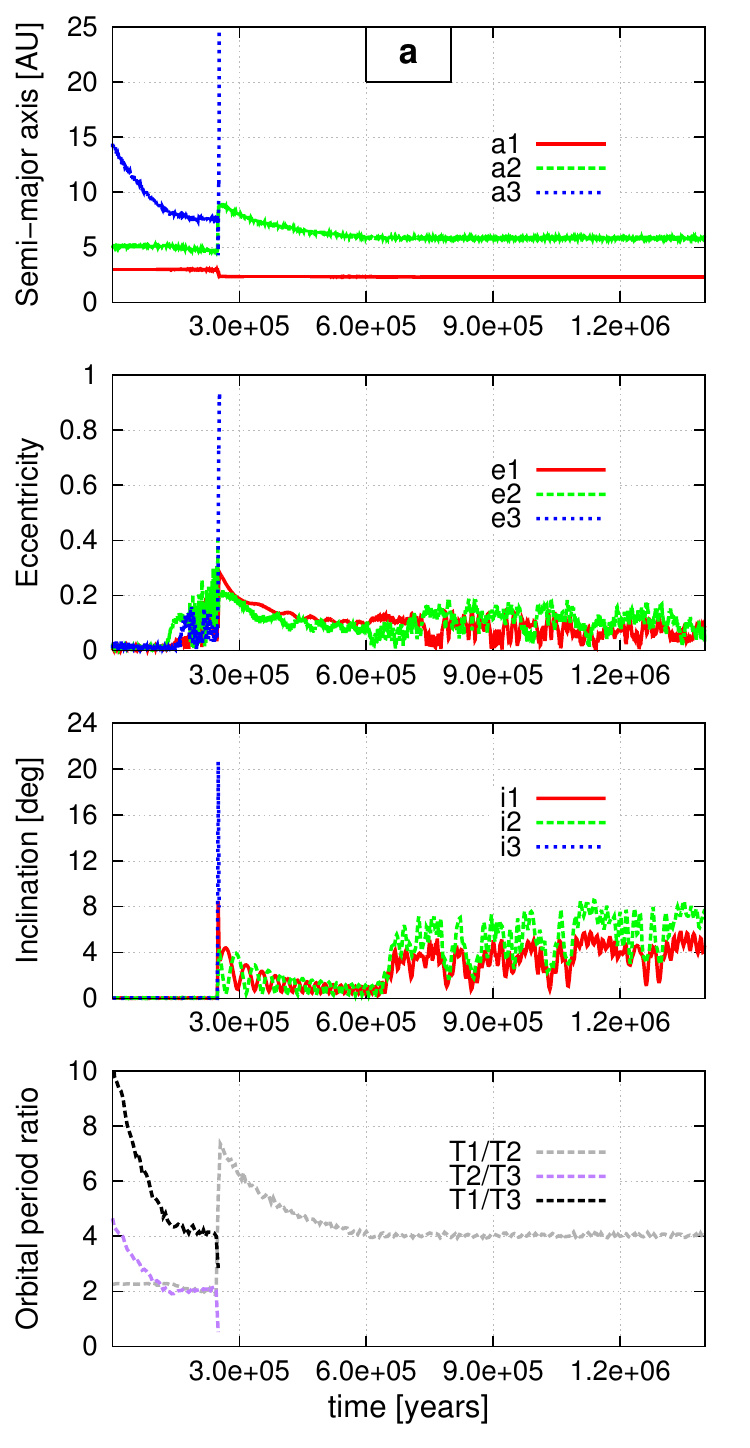}
        \includegraphics[width=0.24\hsize]{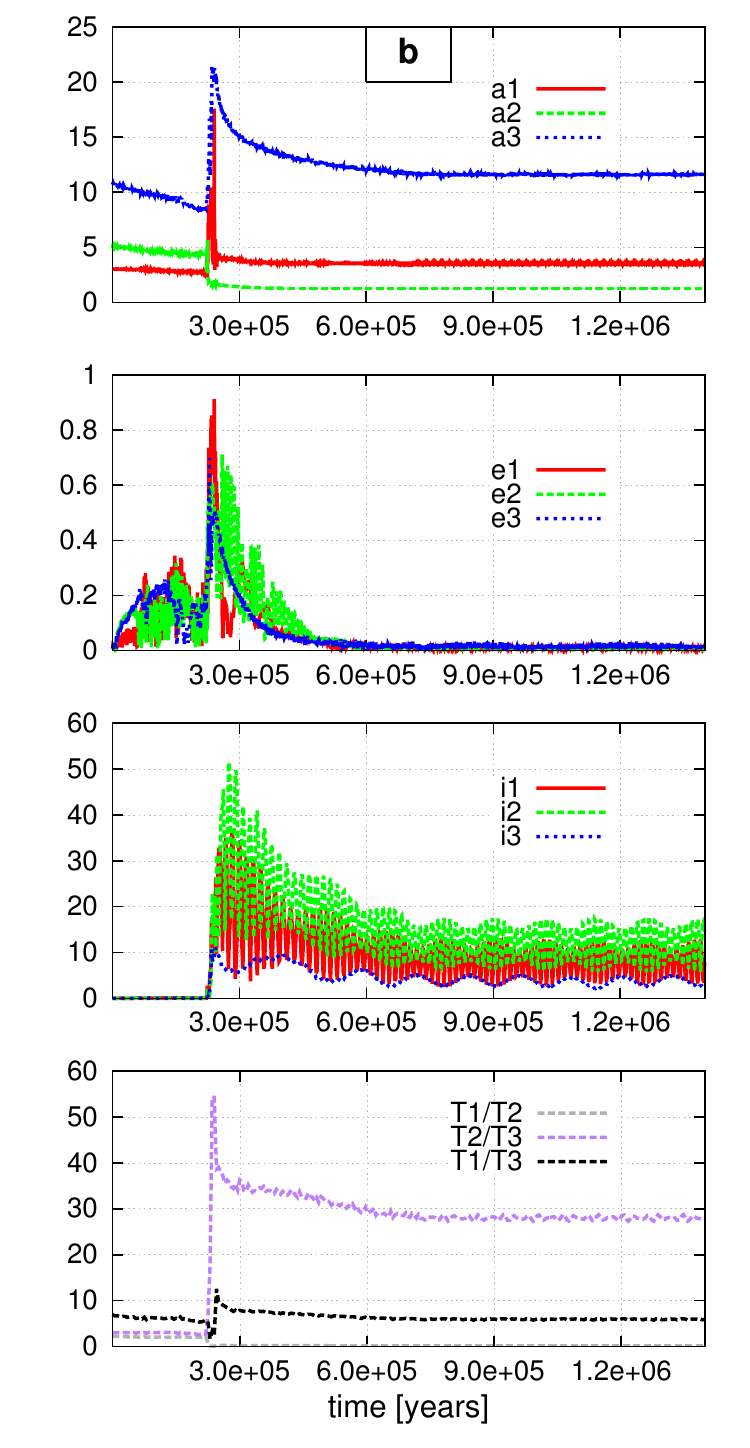}
         \includegraphics[width=0.24\hsize]{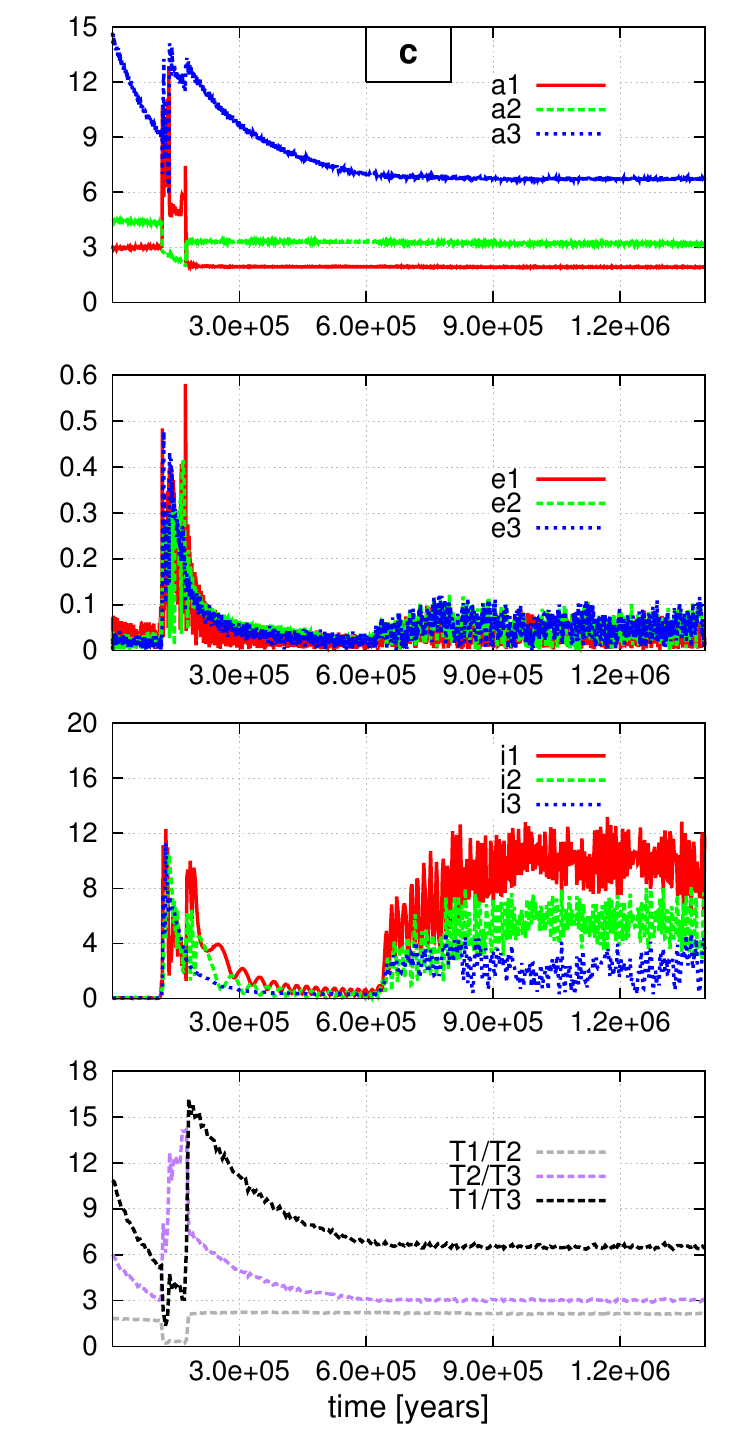}
          \includegraphics[width=0.24\hsize]{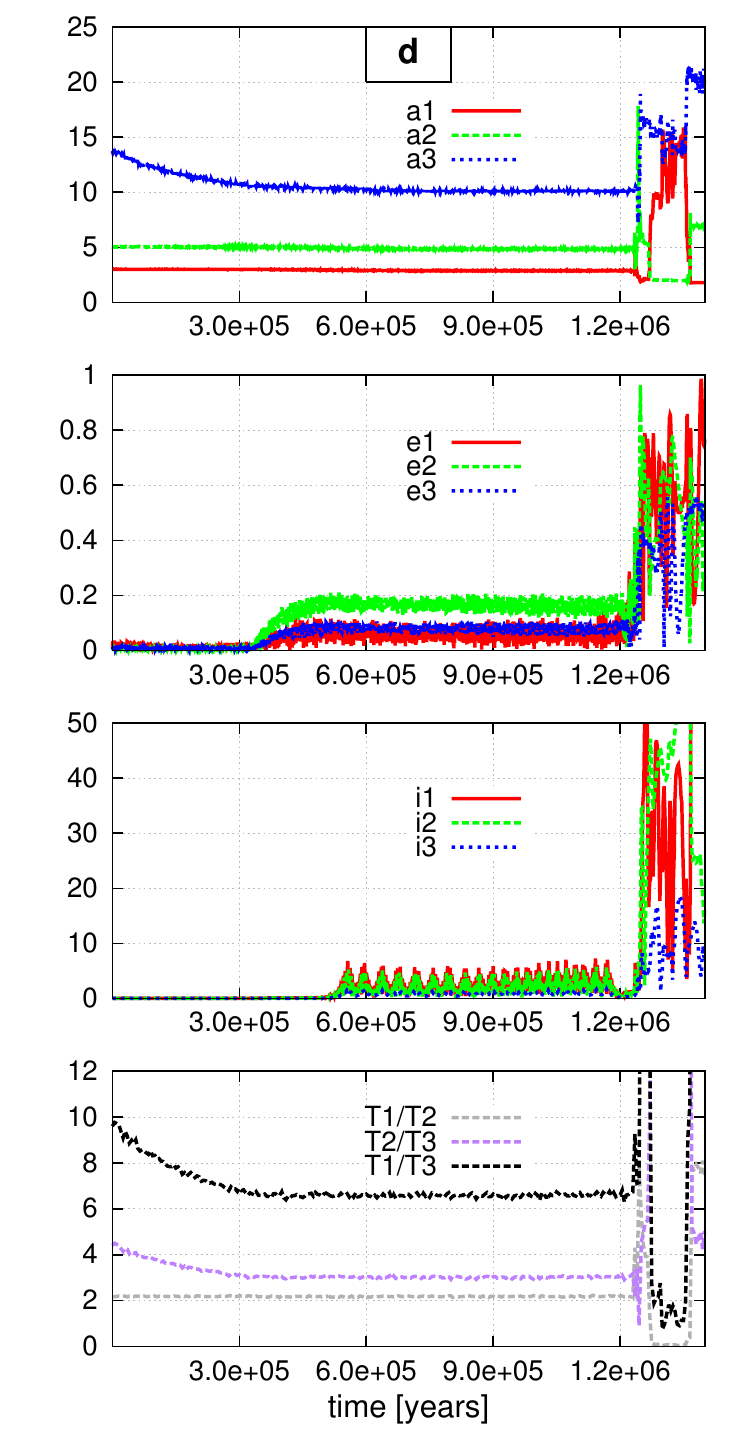}
       \caption{Examples of several scenarios producing mutual inclination excitation: a) planet-planet scattering followed by mean-motion resonance, b) orbital instability, c) orbital instability followed by three-body resonance, and d) three-body resonance followed by orbital instability.}
       \label{fig:scenarios}
 \end{figure*} 
%^^^^^^^^^^^^^^^^^^^^^^^^^^^^^^ %
%^^^^^^^^^^^^^^^^^^^^^^^^^^^^^^ % 

\subsection{Dynamical history of highly mutually inclined systems}\label{4.6}
	Here we focus on systems formed on highly mutually inclined orbits ($\ge 10^\circ$), which represent $\sim\!\!3\%$ of our simulations at the dispersal of the disc. Mechanisms producing inclination increase have been identified in Section \ref{section3}, namely inclination-type resonance and planet-planet scattering. We now aim at understanding the frequency of each mechanism, keeping in mind that the occurrence of both mechanims is possible during the long-term evolution in the disc phase. 

	By carefully studying the dynamical evolution of the highly mutually inclined systems formed in our simulations, we have found seven possible dynamical histories and the frequency of each one is given in Table \ref{tab:history} (third column). Concerning the systems finally composed of two planets at the dispersal of the disc ($1.4 \times 10^6$ yr), three scenarios could have happened: either planet-planet scattering induces the ejection/collision of a planet and excites the mutual inclinations of the remaining planets (similarly to Fig. \ref{fig:fig3}), the inclinations produced by an ejection/collision of a body due to planet-planet scattering are rapidly damped by the disc and a mean-motion resonance capture of the two bodies induces the increase of the inclinations (Fig. \ref{fig:scenarios}, panel a), or a three-body inclination-type resonance excites the inclinations and consequently leads to planet-planet scattering and the ejection of a planet (similarly to Fig. \ref{fig:fig15}). Mutual inclination is always observed in three-body systems following four different histories: either inclination increase is produced by orbital instability of the orbits (without planet ejection) (Fig. \ref{fig:scenarios}, panel b), the orbital elements excited by orbital instability are damped until a three-body resonance capture and a resonant excitation of the inclinations (Fig. \ref{fig:scenarios}, panel c), the system evolves smoothly in a three-body inclination-type resonance (similarly to Fig. \ref{fig:fig4}), or the phase of three-body inclination-type resonance is followed by a destabilization of the orbits (Fig. \ref{fig:scenarios}, panel d).   

	Considering the last mechanism as the source of the inclination excitation, we see that half of the highly mutually inclined systems of our simulations result from two- or three-body mean-motion resonance captures, the other half being produced by orbital instability and/or planet-planet scattering. This emphasizes the importance of mean-motion resonance captures during the disc phase, on the final 3D configurations of planetary systems.
%______________________________________________________________________________________________________________
%%%%%%%%%%%%%%%%%%%%%%%%%%%%%%%%%%%%%%%%%%%%%%%%%%%%%%%%%%%%%%%%%%%%%%%%%%%%%%%%%%%%%%%%%%%%%%%%%%%%%%%%%%%%%%%
%%%%%%%%%%%%%%%%%%%%%%%%%%%%%%%%%%%%%%%%%%%%%%%%%%%%%%%%%%%%%%%%%%%%%%%%%%%%%%%%%%%%%%%%%%%%%%%%%%%%%%%%%%%%%%%
%==============================================================================================================

\section{Long-term evolution}\label{section5}
\begin{figure}%[t!]
       \centering
       \includegraphics[width=0.85\hsize]{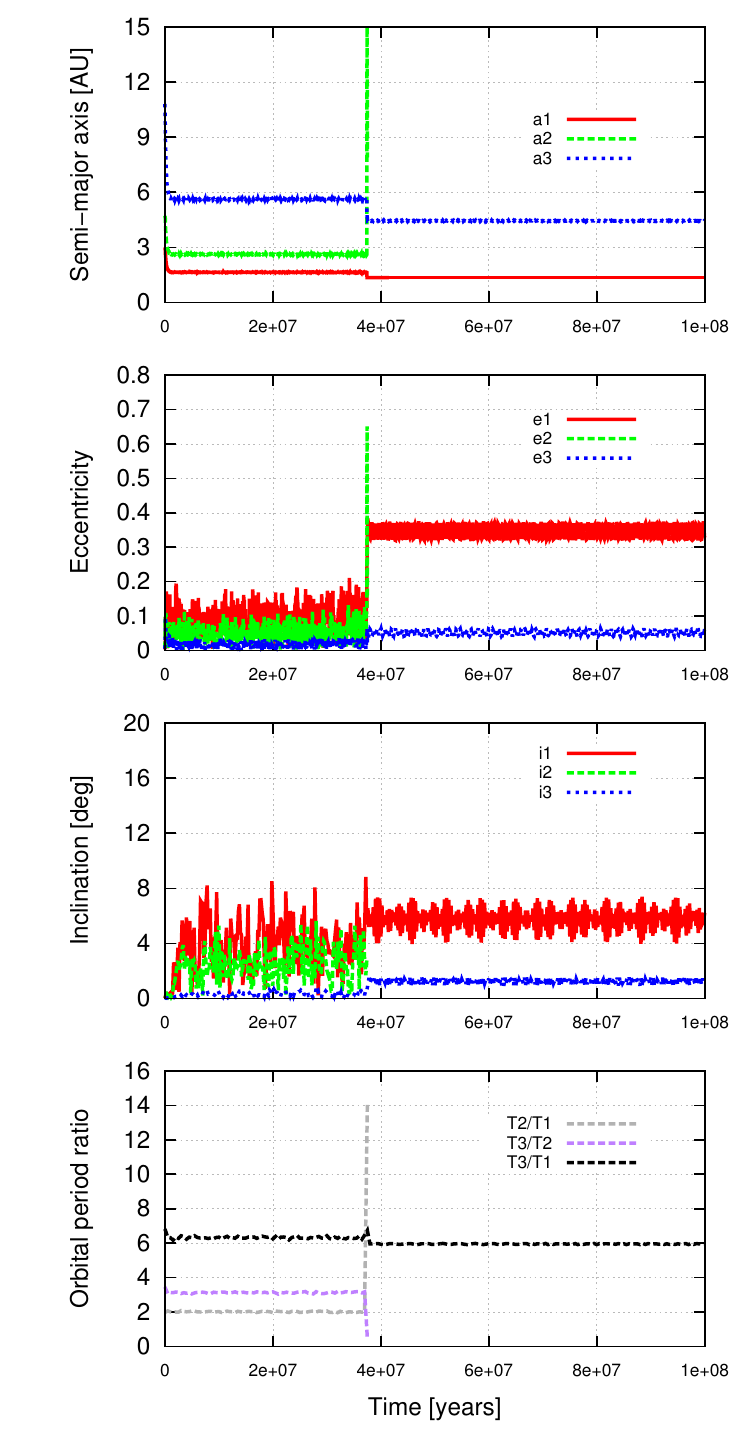} % system index: 3028_371 / M_disc = 16 M_Jup %
       \caption{Destabilization of a three-body resonance on a long timescale. While the system is locked in a $1$:$2$:$6$ resonance at the dispersal of the gas disc, a planet-planet scattering event finally takes place and the middle planet is ejected from the system at $\sim 38$ Myr. The planetary masses are $m_1 = 0.87$, $m_2 = 1.39$ and $m_3 = 8.43$~$M_{\rm Jup}$. The initial mass of the disc is 16 $M_{\rm Jup}$.
 }
       \label{fig:fig15}
 \end{figure}

%==============================================================================================================
%%%%%%%%%%%%%%%%%%%%%%%%%%%%%%%%%%%%%%%%%%%%%%%%%%%%%%%%%%%%%%%%%%%%%%%%%%%%%%%%%%%%%%%%%%%%%%%%%%%%%%%%%%%%%%%
                        %%%%%%%%%%%%%%%%%%%%%%%% Long-term evolution %%%%%%%%%%%%%%%%%%%%%%%%
%%%%%%%%%%%%%%%%%%%%%%%%%%%%%%%%%%%%%%%%%%%%%%%%%%%%%%%%%%%%%%%%%%%%%%%%%%%%%%%%%%%%%%%%%%%%%%%%%%%%%%%%%%%%%%%
        Section \ref{section4} focuses on the distributions of the orbital elements of the planetary systems considered immediately after the dispersal of the disc. As previously discussed, orbital adjustments due to planet-planet interactions can occur on a longer timescale. An example is given in Fig. \ref{fig:fig15}, showing the destabilization of a system in a $1$:$2$:$6$ resonance at the end of the disc phase. At $\sim38$~Myr, the middle planet is ejected from the system and the two surviving bodies are left in well-separated and stable orbits. 

        In this section, we aim to investigate whether or not the long-term evolution of planetary systems produces significant changes on the final distribution of the orbital elements discussed hereabove. To study the long-term evolution of planetary systems, we ran two additional sets of simulations for $100$~Myr: $400$ systems for an initial disc mass of $8$ $M_{\rm Jup}$ and $800$ systems for $16$ $M_{\rm Jup}$, both with an exponential decay of the mass disc (DM model, with the same dispersal time of $\sim\!1$ Myr). As expected, many systems of our long-term simulations are destabilized after millions of years, and the percentages given by Table \ref{tab:title4} on the number of planets in the final configurations have changed significantly. There is a clear tendency towards systems with fewer planets on a long timescale.    

%^^^^^^^^^^^^^^^^^^^^^^^^^^^^^^ %
%^^^^^^^^^^ Figure 15 ^^^^^^^^^ %
%^^^^^^^^^^^^^^^^^^^^^^^^^^^^^^ %
%^^^^^^^^^^^^^^^^^^^^^^^^^^^^^^ %
%^^^^^^^^^^^^^^^^^^^^^^^^^^^^^^ %  

%^^^^^^^^^^^^^^^^^^^^^^^^^^^^^^ %
% ^^^^^^^^^ Figure 17 ^^^^^^^^^ %
%^^^^^^^^^^^^^^^^^^^^^^^^^^^^^^ %
 \begin{figure}%[t!]
       \centering
       \includegraphics[width=\hsize]{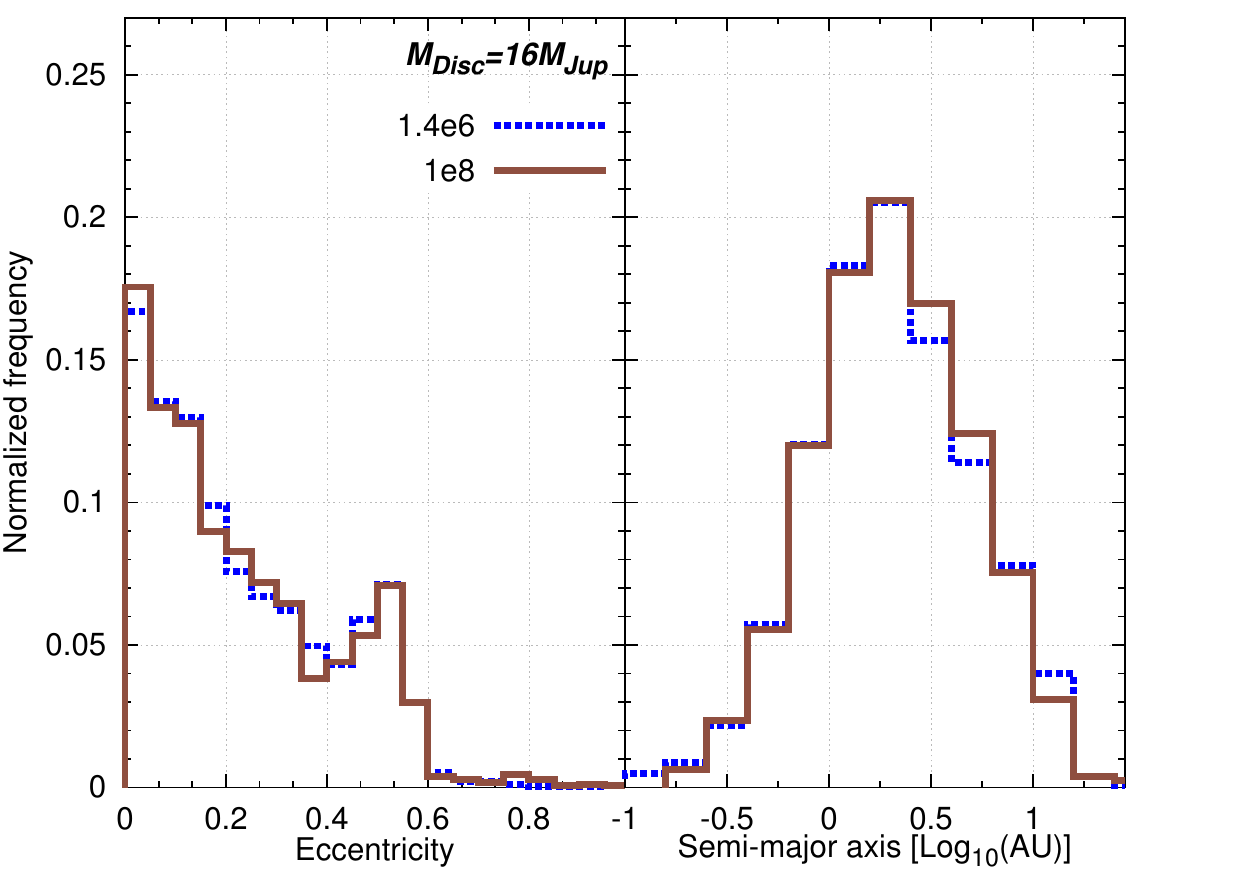}  % Multi-figure for semi-major axis distribution
       \caption{{\it Left panel:} Normalized eccentricity distributions of the $16$ $M_{\rm Jup}$ disc simulations for two integration timescales; $1.4\times10^6$ and $1\times10^8$ yr. The bin size is $\Delta e = 0.05$. {\it Right panel:} Normalized semi-major axis distributions of the $16$ $M_{\rm Jup}$ disc simulations, for the same two integration timescales. The bin size is $\Delta \log(a) = 0.2$.}
       \label{fig:fig17}
 \end{figure}
%^^^^^^^^^^^^^^^^^^^^^^^^^^^^^^ %
%^^^^^^^^^^^^^^^^^^^^^^^^^^^^^^ %

%^^^^^^^^^^^^^^^^^^^^^^^^^^^^^^ %
%^^^^^^^^^^ Figure 18 ^^^^^^^^^ %
%^^^^^^^^^^^^^^^^^^^^^^^^^^^^^^ %
 \begin{figure}%[t!]
       \centering
       \includegraphics[width=\hsize]{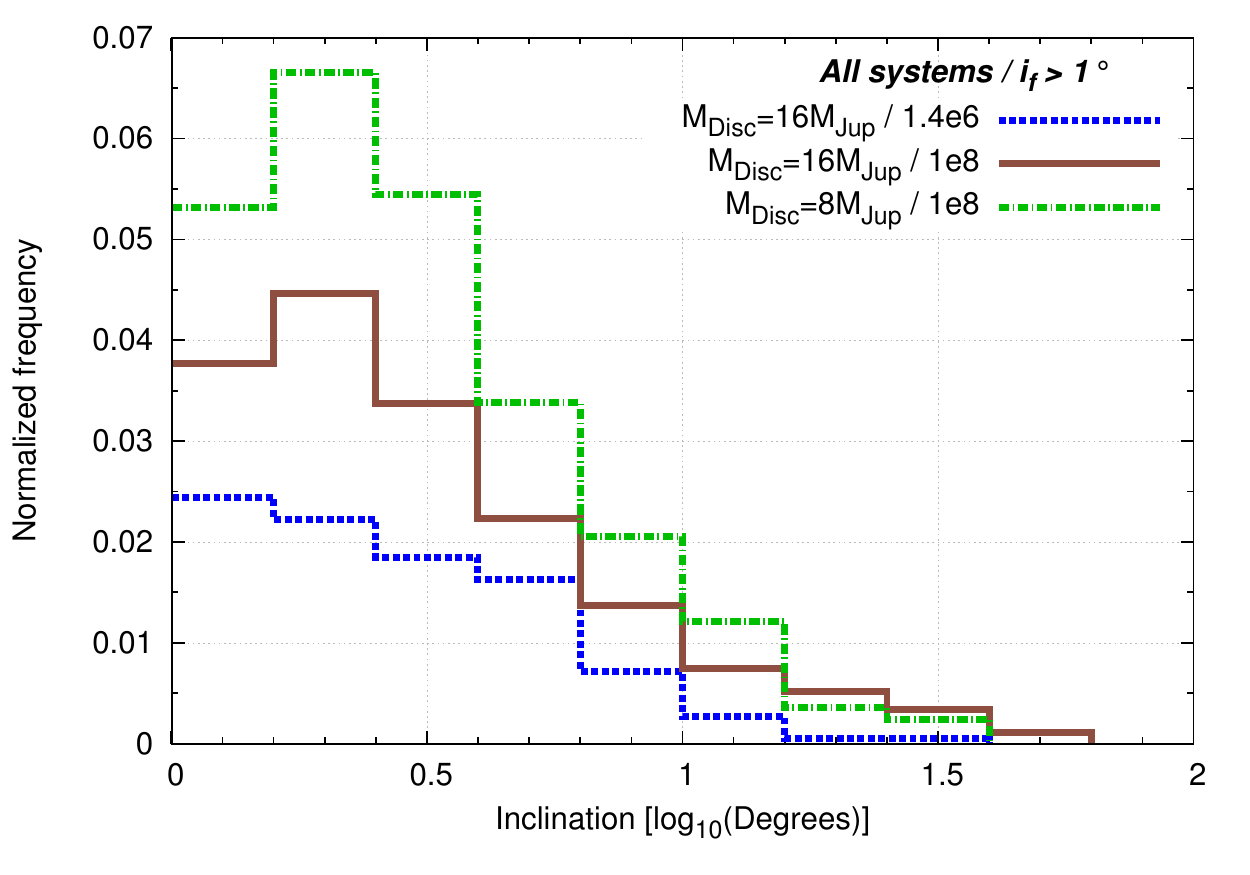}  % Multi-figure for semi-major axis distribution
       \caption{Normalized inclination distributions for two integration timescales ($1.4\times10^6$ and $1\times10^8$ yr) and two initial disc masses ($8$ and $16$ $M_{\rm Jup}$). Only the inclinations higher than $1^{\circ}$ are represented. The bin size is $\Delta \log(i) = 0.2$.}
       \label{fig:fig18}
 \end{figure}
%^^^^^^^^^^^^^^^^^^^^^^^^^^^^^^ %
%^^^^^^^^^^^^^^^^^^^^^^^^^^^^^^ %

        Fig. \ref{fig:fig17} shows no significant change on the semi-major axis and eccentricity distributions on the $100$ Myr timescale. However, the inclinations have considerably increased on a longer timescale, as appears clearly in Fig. \ref{fig:fig18}. Just after the disc phase, $\sim 10\%$ of the planets have inclinations higher than $1^{\circ}$ and this percentage has almost doubled at $100$ Myr ($\sim 25\%$ for the $8$~$M_{\rm Jup}$ and $\sim 17\%$ for the $16$ $M_{\rm Jup}$). 

        Concerning the mutual inclinations, $\sim 3\%$ of the three-planet systems are highly mutually inclined ($I_{mut} > 10^{\circ}$) at $1.4\times10^6$~yr and this percentage remains approximately the same for the long-term simulations. The situation is quite different for the two-planet systems. There are approximately $2\%$ of highly mutually inclined systems at the dispersal of the gas disc and $\sim 7\%$ at $100$ Myr. As a result, $\sim 5\%$ of the multiple systems of our simulations have high mutual inclinations ($I_{mut} > 10^{\circ}$). 
  
        The destabilization of highly mutually inclined systems is also common when considering the long-term evolution on $100$ Myr. The same observation has recently been pointed out by \citet{Barnes2015}, concerning planetary systems in mean-motion resonance with mutual inclinations. Referring to Table \ref{tab:history} (fourth column), we see that the percentage of highly mutually inclined systems still evolving in resonance drops to $30\%$ (instead of $50\%$ at the dispersal of the disc).
%______________________________________________________________________________________________________________
%%%%%%%%%%%%%%%%%%%%%%%%%%%%%%%%%%%%%%%%%%%%%%%%%%%%%%%%%%%%%%%%%%%%%%%%%%%%%%%%%%%%%%%%%%%%%%%%%%%%%%%%%%%%%%%
%%%%%%%%%%%%%%%%%%%%%%%%%%%%%%%%%%%%%%%%%%%%%%%%%%%%%%%%%%%%%%%%%%%%%%%%%%%%%%%%%%%%%%%%%%%%%%%%%%%%%%%%%%%%%%%
%==============================================================================================================

\section{Conclusions}\label{section6}
%==============================================================================================================
%%%%%%%%%%%%%%%%%%%%%%%%%%%%%%%%%%%%%%%%%%%%%%%%%%%%%%%%%%%%%%%%%%%%%%%%%%%%%%%%%%%%%%%%%%%%%%%%%%%%%%%%%%%%%%%
                                                        %%%%%%%%%%%%%%%%%%%%% Conclusions %%%%%%%%%%%%%%%%%%%%%
%%%%%%%%%%%%%%%%%%%%%%%%%%%%%%%%%%%%%%%%%%%%%%%%%%%%%%%%%%%%%%%%%%%%%%%%%%%%%%%%%%%%%%%%%%%%%%%%%%%%%%%%%%%%%%%
        In this study we followed the orbital evolution of three giant planets in the late stage of the gas disc. Our scenario for the formation of planetary systems combines Type-II migration, with the consistent eccentricity and inclination damping of Paper I \citep{Bitsch2013}, and planet-planet scattering. The results shown in this work are naturally attached to the disc parameters of the hydrodynamical simulations of Paper I from which the damping formulae are issued. Our parametric experiments consisted of 11000 numerical simulations, considering a variety of initial configurations, planet mass ratios, and disc masses. Moreover, two modelizations of the gas disc were taken into account: the constant-mass model (no gas dissipation during $0.8$ Myr) and the decreasing-mass model (gas exponential decay with an e-folding timescale of $1$ Myr). The first case leads to more merging, more migration, and less ejections of planets.  
        
        We focused on the impact of eccentricity and inclination damping on the final configuration of planetary systems. We have shown that the eccentricities are already well-diversified at the dispersal of the disc, despite the strong eccentricity damping exerted by the gas disc, and subsequent inclination increase is possible. Concerning the inclinations, in contrast with previous works that did not include inclination damping, we found that most of the planets end up in the midplane of the disc (i.e., in quasi-coplanar orbits with $i < 1^{\circ}$), showing the efficiency of the inclination damping. One should keep in mind that planets formed in the disc midplane could appear inclined with respect to the stellar equatorial plane if these two planes differ. Needless to say, the higher the initial disc mass, the smaller the inclinations of the planets at the dispersal of the disc. Nevertheless, in multiple systems, inclination-type resonance and planet-planet scattering events during/after the gas phase can produce inclination excitation of the inclinations. Approximately $5\%$ of highly mutually inclined systems ($I_{mut}>10^{\circ}$) have been formed in our scenario. In future observations, this percentage could help to discriminate between the formation scenarios.  

        The dynamical mechanisms producing inclination increase were identified, namely inclination-type resonance, orbital instability, and/or planet-planet scattering. We have shown that resonance captures play an important role in the formation of mutually inclined systems. Indeed, half of these systems originate from two- or three-body mean-motion resonance captures. The long-term evolution of the systems was been investigated, showing that destabilization of the resonant systems is common. However, $~30\%$ of the systems still evolve in resonance after $100$ Myr.  

        As a by-product of our study, we found a very good agreement between our simulations and the observed population of extrasolar systems, in particular for the semi-major axis and eccentricity distributions. Although a full exploration of the parameter space and a real population synthesis study are far beyond the scope of this paper, this agreement suggests strongly that planet-planet interactions during the migration inside the protoplanetary disc could account for most of the eccentricity excitation observed among exoplanets.  
%______________________________________________________________________________________________________________
%%%%%%%%%%%%%%%%%%%%%%%%%%%%%%%%%%%%%%%%%%%%%%%%%%%%%%%%%%%%%%%%%%%%%%%%%%%%%%%%%%%%%%%%%%%%%%%%%%%%%%%%%%%%%%%
%%%%%%%%%%%%%%%%%%%%%%%%%%%%%%%%%%%%%%%%%%%%%%%%%%%%%%%%%%%%%%%%%%%%%%%%%%%%%%%%%%%%%%%%%%%%%%%%%%%%%%%%%%%%%%%
%==============================================================================================================

%==============================================================================================================
%%%%%%%%%%%%%%%%%%%%%%%%%%%%%%%%%%%%%%%%%%%%%%%%%%%%%%%%%%%%%%%%%%%%%%%%%%%%%%%%%%%%%%%%%%%%%%%%%%%%%%%%%%%%%%%
                                                        %%%%%%%%%%%%%%%%%%%%% Acknowledgments %%%%%%%%%%%%%%%%%%%%%
%%%%%%%%%%%%%%%%%%%%%%%%%%%%%%%%%%%%%%%%%%%%%%%%%%%%%%%%%%%%%%%%%%%%%%%%%%%%%%%%%%%%%%%%%%%%%%%%%%%%%%%%%%%%%%%
\begin{acknowledgements}
This work was supported by the Fonds de la Recherche Scientifique-FNRS under Grant No. T.0029.13 ("ExtraOrDynHa" research project). Computational resources were provided by the Consortium des \'Equipements de Calcul Intensif (C\'ECI), funded by the Fonds de la Recherche Scientifique de Belgique (F.R.S.-FNRS) under Grant No. 2.5020.11. We thank the referee for his constructive suggestions that have improved the manuscript.  
\end{acknowledgements}
%______________________________________________________________________________________________________________
%%%%%%%%%%%%%%%%%%%%%%%%%%%%%%%%%%%%%%%%%%%%%%%%%%%%%%%%%%%%%%%%%%%%%%%%%%%%%%%%%%%%%%%%%%%%%%%%%%%%%%%%%%%%%%%
%%%%%%%%%%%%%%%%%%%%%%%%%%%%%%%%%%%%%%%%%%%%%%%%%%%%%%%%%%%%%%%%%%%%%%%%%%%%%%%%%%%%%%%%%%%%%%%%%%%%%%%%%%%%%%%
%==============================================================================================================

%==============================================================================================================
%%%%%%%%%%%%%%%%%%%%%%%%%%%%%%%%%%%%%%%%%%%%%%%%%%%%%%%%%%%%%%%%%%%%%%%%%%%%%%%%%%%%%%%%%%%%%%%%%%%%%%%%%%%%%%%
                                                %%%%%%%%%%%%%%%%%%%%% Bibliography %%%%%%%%%%%%%%%%%%%%%
%%%%%%%%%%%%%%%%%%%%%%%%%%%%%%%%%%%%%%%%%%%%%%%%%%%%%%%%%%%%%%%%%%%%%%%%%%%%%%%%%%%%%%%%%%%%%%%%%%%%%%%%%%%%%%%
\bibliographystyle{aa}
\bibliography{bibliography}


\end{document}